\documentclass[12pt]{iopart}

\usepackage{iopams}  
\usepackage{graphicx}
\usepackage{dcolumn}
\usepackage{longtable}
\usepackage{tabularx}
\usepackage{amssymb}
\usepackage{color}
\begin{document}

\title[]
{The finite range simple effective interaction including tensor terms}  
\author{P. Bano$^\oplus$}
\address{School of Physics, Sambalpur University, Jyotivihar-768 019, India.}
\author{X. Vi\~nas$^{\ddag}$}
\address{Departament de F\'isica Qu\`antica i Astrof\'isica and Institut de
Ci\`encies del Cosmos (ICCUB), Facultat de F\'isica, Universitat de Barcelona, Mart\'i
i Franqu\`es 1, E-08028 Barcelona, Spain}
\author{T. R. Routray$^{\dagger}$}
\address{School of Physics, Sambalpur University, Jyotivihar-768 019, India.}
\author{M. Centelles$^{\parallel}$}
\address{Departament de F\'isica Qu\`antica i Astrof\'isica and Institut de
Ci\`encies del Cosmos (ICCUB), Facultat de F\'isica, Universitat de Barcelona, Mart\'i
i Franqu\`es 1, E-08028 Barcelona, Spain}
\author{M. Anguiano$^{\S}$}
\address{Departamento de F\'{\i}sica At\'omica, Molecular y Nuclear,
Universidad de Granada, E-18071 Granada, Spain}
\author{L. M. Robledo$^{*}$}
\address{Universidad Aut\'onoma de Madrid,E-28049 Madrid, Spain.}
\address{Center for Computational Simulation, Universidad Polit\'ecnica de Madrid,
Campus de Montegancedo, Boadilla del Monte, 28660-Madrid, Spain.}

\ead{$^{\ddag}$xavier@fqa.ub.edu,$^{\dagger}$trr1@rediffmail.com(Corresponding author), $^{\parallel}$mariocentelles@ub.edu, $^{\S}$mangui@ugr.es, $^{*}$luis.robledo@uam.es, $^\oplus$mailme7parveen@gmail.com}
\vspace{10pt}

\begin{abstract}
The prediction of single particle level crossing phenomenon between $2p_{3/2}$ and $1f_{5/2}$ orbitals in $Ni$- 
and $Cu$-isotopic chains by the finite range simple effective interaction without requiring the tensor part is discussed. 
In this case the experimentally observed crossing could be studied as a function of nuclear matter incompressibility, $K(\rho_0)$. 
The estimated crossing for the neutron number $N$=46 could be reproduced by the equation of state corresponding to $K(\rho_0)$=240 MeV. 
However, the observed proton gaps between the $1h_{11/2}$ and $1g_{7/2}$ shells in $Sn$ and $Sb$ isotopic chain, 
and the neutron gaps between the $1i_{13/2}$ and $1h_{9/2}$ shells in $N$=82 isotones, as well as the shell closure properties at $N$=28 
require explicit consideration of a tensor part as the central contribution is not enough to initiate the required level splittings. 
\end{abstract}

%
%
%
%
%

\section*{Introduction}
Different characteristics features contained in the nucleon-nucleon (NN) interaction on the top of the central short-range attractive 
part manifest at situations to give rise to observable effects. With the progress in the facility of producing neutron-rich isotopes far from 
beta-stability limit, the relevance of the tensor component in the NN interaction has gained importance over the last few decades. Since long ago 
the signature of a tensor component contained in the nuclear force was known from the non-vanishing electric quadrupole moment in deuteron 
\cite{Rarita1941,Blatt1952} but its importance in the nuclear studies using Skyrme or Gogny mean-field models have been undermined over several 
decades for its small contribution to the bulk nuclear properties. But the theoretical interpretation of the experimental data 
\cite{ Schiffer2004,Gaudefroy2006,Colo2007} on the energy gaps between single particle (s.p.) levels arising from the isotopic and isotonic effects 
once again established the crucial relevance of the tensor force. 
The presence of a tensor term in the interaction is mandatory to study 
new phenomena related to the evolution of the spin-orbit splittings with the neutron excess in exotic neutron-rich nuclei, as it has been pointed
out, for instance, in  Refs. \cite{Otsuka2005,Otsuka2010,Otsuka2001}.

The crossing of the $2p_{3/2}$ and $1f_{5/2}$ proton s.p. energy levels in neutron-rich $Ni$ isotopes   
and the magic character of the atomic number $Z$=28 in this isotopic chain is a subject of current interest from both, experimental 
and theoretical points of view \cite{Olivier,Sahin}. The dominant s.p. character of some of the low lying excited states in neutron-rich 
$Cu$ isotopes, ascertained from the $\gamma$-spectroscopic studies \cite{Sahin}, justifies the analysis of the data using mean-field model 
calculation. Some phenomenological effective NN forces of Skyrme and Gogny types \cite{Brink2018,Anguiano2012}, as well as the 
microscopic calculations based on meson-exchange interactions \cite{Otsuka2005}, are able to describe theoretically the $1f_{5/2}$ and $2p_{3/2}$ s.p. 
proton levels crossing in $Ni$-isotopes if a tensor component is included in the NN interaction. The underlying reason is that the tensor force 
modifies the spin-orbit (SO) splitting of the s.p. levels, which is crucial for obtaining the experimentally observed crossing in neutron-rich 
$Ni$ isotopes. 
Using the Skyrme interactions Sk-III and SAMi-T, this crossing in $Ni$-isotopes has been achieved by adding a short-range momentum 
dependent tensor part \cite{Brink2018,Shen2019}. On the other hand, Anguiano et al have obtained the 
aforementioned crossing at $A$=74 by performing spherical Hartree-Fock (HF) calculations in coordinate space using the D1M 
Gogny interaction together with a finite-range tensor contribution \cite{Anguiano2012,Anguiano2016,Anguiano2011}. This crossing predicted by the 
D1M set can also be obtained in the framework of quasi-local density functional theory (QLDFT) \cite{Soubbotin2000,Soubbotin2003} by adding a 
short-range tensor part as the one used in Skyrme forces, as has been shown in a recent work \cite{Routray2021}. 
In this reference it has also been shown that the finite range simple 
effective interaction (SEI) predicts the crossing of the $1f_{5/2}$ and $2p_{3/2}$ s.p. levels in $Ni$-isotopes without requiring any explicit 
tensor term. This fact points out the relevance of the underlying central mean-field. In the case of SEI this 
level crossing phenomenon can be studied as a function of nuclear matter (NM) parameter, in particular, the incompressibility $K(\rho_0)$, as will be discussed 
in the present work. However, there are many other phenomena discussed in Refs. \cite{Schiffer2004,Otsuka2005} for which explicit consideration 
of a tensor term together with the SEI becomes essential to obtain the experimental trend. 
Our main aim in this work is to enlarge the simple effective interaction model by including a tensor term, which is chosen as short-range,
similar to that used with Skyrme forces \cite{Brink2018}. The two strength parameters of the tensor term are fixed subject to the constraint that 
the predicted crossing of $1f_{5/2}$ and $2p_{3/2}$ proton s.p. levels in $Ni$-isotopes and the spin inversion in the ground-state of
$Cu$-isotopes remain unchanged.
The paper is organized as follows. In the second section we revise the basic aspects of the mean-field approach based on the SEI paying special
attention to the fitting protocol of its parameters, which is somewhat different from the one used in other effective interactions of Skyrme or
Gogny types. In the same section, the short-range tensor force is briefly discussed. The third section is devoted to the discussion of the
results obtained in this work. 
First, we show the impact of the nuclear mean-field on the proton level crossing in Ni isotopes. Next, we analyze the splitting of the 
$1h_{11/2}-1g_{7/2}$ s.p. proton levels along the $Sn$ and $Sb$ isotopic chains and the $1i_{13/2}-1h_{9/2}$ splitting of the s.p. neutron level in 
isotones of N=82 with the SEI model including a tensor contribution. Finally, using this model, we discuss the reduction of the spin-orbit 
splittings of the $1f$ and $2p$ energy levels along the $N$=28 isotonic chains when the neutron-proton asymmetry increases. Finally, 
our conclusions are given in the last section.

\section*{Formalism}

\subsection*{The simple effective interaction}

 The SEI was introduced in Ref.\cite{Behera1998} by Behera and collaborators. It contains altogether 11-numbers of 
parameters in nuclear matter, namely, $\alpha$, $\gamma$, $b$, $x_{0}$, $x_{3}$, 
$t_{0}$, $t_{3}$, $W$, $B$, $H$, and $\ M$. This interaction reads: 
\begin{eqnarray}\label{1}
	  V_{eff}&=&t_{0}(1+x_{0}P_{\sigma})\delta(\vec{r}) +\frac{t_{3}}{6}(1+x_{3}P_{\sigma})\left(\frac{\rho(\vec{R})}
{1+b\rho(\vec{R})}\right)^{\gamma}\delta(\vec{r})\nonumber \\
	&& +(W+BP_{\sigma}-HP_{\tau}-MP_{\sigma}P_{\tau})f(\vec{r}) + Spin-orbit part.
\label{SEI}
 \end{eqnarray}
An additional parameter, namely the strength of the the spin-orbit interaction $W_0$, comes into the picture 
in the description of finite nuclei. The protocol adopted for the parameter determination in case of SEI is somewhat different from that of adopted 
in case of other effective forces. The parameter combinations responsible for the momentum dependence of the mean fields in nuclear matter of 
different types are fixed from experimental/empirical conditions. In this context, we impose that the mean field in symmetric 
nuclear matter (SNM) changes sign at kinetic energy 300 MeV \cite{ber1988,gale87,gale90,cser92}, the entropy in pure neutron matter (PNM) should not 
exceed that of SNM \cite{Behera11}, and the effective mass splitting in spin polarized neutron matter compares to that of the 
Dirac-Bruckner-Hartree-Fock (DBHF) prediction \cite{Behera2015}. The parameter combinations responsible for the density dependence of the isospin 
asymmetric part of the equation of state (EoS) are fixed from the standard value of the saturation properties alongwith the empirical condition that 
the asymmetric nucleonic contribution in $\beta$-stable charge neutral $n+p+e+\mu$ matter be maximum. The stiffness of the SNM EOS is determined by 
the exponent $\gamma$ of the density dependent $t_3$-term, which  
is kept as a free parameter and all values of $\gamma$ for which the pressure-density relation remains within the allowed range extracted from the 
analysis of the heavy-ion collision data at intermediate energies \cite{Danielewicz2002} are allowed. The upper limit of $\gamma\approx1$  thus 
obtained for the Gaussian form of SEI corresponds to the value of incompressibility $K$=283 MeV. The density dependent term is also modified with 
a denominator containing the parameter $b$, which is ascertained so as to prevent the NM to become supraluminous \cite{Behera1997}. With the 
parameters determined in this way, the SEI was able to reproduce the trends of the EoS and the momentum dependence of the mean field properties in NM 
with similar quality as predicted by microscopic calculations \cite{Behera11,Sammarruca2010,Wiringa1988,APR1998,Behera09}.
For finite nuclei calculation only remains one NM parameter open, which has been taken as $t_{0}$. 
This parameter together with the SO-strength $W_0$ are fixed by fitting the binding energies (BEs) of the magic nuclei 
$^{40}$Ca and $^{208}$Pb.  
In our study, finite nuclei are described through the so-called Quasi-local Density Functional Theory (QLDFT). It is a HF calculation 
performed starting from the quantal energy density, but with the exchange contribution localized thorough the extended Thomas-Fermi approximation
of the one-body density matrix, which contains up to second order terms in the $\hbar$-expansion \cite{Soubbotin2000,Soubbotin2003}. 
The corresponding Schrodinger equation results into a set of coupled single-particle equations, 
\begin{equation}
\left[- \nabla.\frac{\hbar^{2}}{2m*_{q}}\nabla+U_{q}(\vec{R})-\vec{W}_q(\vec{R})(\nabla \times \vec{\sigma})\right]\phi_{q}=\epsilon_{q}\phi_{q}, 
\label{eq1}
\end{equation}
where the subscript $q=n,p$ indicates the type of particle, $m^{*}_{q}$ is the effective mass, $U_{q}$ is the mean-field experienced by the 
nucleon $q$, $\vec{W}_q$ is the form-factor of the spin-orbit potential and $\epsilon_{q}$ is the s.p. energy corresponding to 
the orbital $\phi_{q}$. This set of equations, which are local in the 
coordinate space, can be solved in the case of spherical symmetry in a similar way to that used for Skyrme forces.
The excellent  agreement between the predictions  for spherical nuclei obtained using the QLDFT approximation and the full HF results
has been discussed in detail in Ref.\cite{Behera2016}. Just as an example connected with our present study, we report in Table \ref{Ni_BE}
the BEs of $Ni$-isotopes from $A$=68 to 78 computed at QLDFT level for the four EoSs of SEI, corresponding to the $\gamma$-values $\frac{1}{6}$, 
$\frac{1}{3}$, $\frac{1}{2}$, and $\frac{2}{3}$, alongwith the experimental values. To deal with the pairing correlations in open-shell nuclei, 
we use the BCS approach together with a zero-range 
density-dependent pairing interaction of the type proposed by Bertsch and Esbensen and  whose parameters were fitted to reproduce the pairing gaps 
in NM predicted by the Gogny interactions (see \cite{Behera2016,Behera2013} for more details.) 
\begin{table}
	\caption{$Ni$ nuclei ground state energy for $A$=68 to 78 calculated for the four EoS of SEI compared with experimental value\cite{nndc}.}
	\begin{center}
		\begin{tabular}{cccccc}
			\hline
			Nuclei&SEI($\gamma=\frac{1}{6}$)&SEI($\gamma=\frac{1}{3}$)&SEI($\gamma=\frac{1}{2}$)&SEI($\gamma=\frac{2}{3}$)&Expt.\cite{nndc}\\
			&E[MeV]&E[MeV]&E[MeV]&E[MeV]&E[MeV]\\
			\hline
			\hline
			$^{68}$Ni&-591.60&-591.08&-590.37&-590.46&-590.407\\
			$^{70}$Ni&-604.76&-604.52&-603.80&-603.82&-602.300\\
			$^{72}$Ni&-616.44&-616.32&-615.73&-615.83&-613.455\\
			$^{74}$Ni&-627.04&-627.03&-626.49&-626.71&-623.82\\
			$^{76}$Ni&-636.64&-636.75&-636.27&-636.53&-633.156\\
			$^{78}$Ni&-645.81&-645.38&-644.96&-645.27&-641.55\\
			\hline
			\label{Ni_BE}	
		\end{tabular}	
	\end{center}
\end{table}
%
Several studies in the domain of nuclear matter under extreme conditions Ref.\cite{Behera11,Behera09,Routray16} as well as in finite nuclei 
Ref.\cite{Behera2015,Behera2016,Behera2013}, have been made using EoSs of SEI corresponding to the $\gamma$-values $\frac{1}{6}$, $\frac{1}{3}$, 
$\frac{1}{2}$, and $\frac{2}{3}$. These EoSs for the $\gamma$-values correspond to incompressibility in SNM 207, 226, 245, and 263 MeV, respectively.
In this work we will also use the SEI EoS with $\gamma$=0.42, whose parameters alongwith the saturation properties are reported below.
\subsection*{The spin-orbit and the tensor force}
The Skyrme-type spin-orbit interaction is used whose contribution to the energy density is given by,
\begin{eqnarray}
&&{\cal H}_{SO} = -\frac{W_o}{2} \big[\rho\nabla\cdot{\bf J}+ \rho_n\nabla\cdot{\bf J_n}+\rho_p\nabla\cdot{\bf J_p}\big].
\label{Eq_H_SO}
\end{eqnarray}
The spin orbit densities $J_q$ ($q=n,p$) are given by
\begin{eqnarray}
&&{\bf J}_q({\bf r}) = \frac{1}{4\pi r^3}\sum_{i} v^{2}_{i}(2j_i+1)\big[j_i(j_i+1)- l_i(l_i+1)-\frac{3}{4}\big]R^{2}_i(r),
\label{Eq_J}
\end{eqnarray}
 where the sum index $i$ runs over all the quantum number labeled by $i=n, l, j$, 
$R_i$ is the radial part of the wave function and $v_i$ is the BCS occupation probability of the state.  
The contribution to the SO potential is obtained from the variation of $H_{SO}$ with respect to $J_q$, $q=n,p$, which results into 
\begin{eqnarray}
&&{\bf W}_q = \frac{W_o}{2} \left[2\nabla\rho_q+ \nabla\rho_{q'}\right].
\label{Eq_V_SO}
\end{eqnarray}
In this work we enlarge SEI by adding a tensor term, which is taken as a short-range force as the one used in Skyrme 
interactions. We have checked previously \cite{Routray2021} that a QLDFT calculation with the Gogny D1M force together with a short-range 
tensor predicts a finite nuclei description extremely close to the one obtained in full HF calculations with a finite-range tensor 
\cite{Anguiano2012,Anguiano2016,Anguiano2011}. Although the contribution of the zero-range tensor force to the energy density functional has been
discussed in detail in earlier literature \cite{Colo2007,Brink2018}, we briefly summarize it here for a sake of completeness.

The short-range tensor force including triplet-even and triplet-odd terms, with strengths $T$ and $U$ respectively, reads:
\begin{eqnarray}
&&V_T = \frac{T}{2}\bigg\{\bigg[(\mathbf{\sigma_1}\cdot{\bf k'})(\mathbf{\sigma_2}\cdot{\bf k'})
-\frac{1}{3}(\mathbf{\sigma_1}\cdot\mathbf{\sigma_2)}{\bf k'}^2\bigg] \delta({\bf r_1 - r_2}) \nonumber \\ 
&& +\delta({\bf r_1 - r_2})\bigg[(\mathbf{\sigma_1}\cdot{\bf k})(\mathbf{\sigma_2}\cdot{\bf k})
-\frac{1}{3}(\sigma_1\cdot\sigma_2){\bf k}^2\bigg]\bigg\} \nonumber \\
&&+U\bigg\{(\mathbf{\sigma_1}\cdot{\bf k'})\delta({\bf r_1 - r_2})(\mathbf{\sigma_2}\cdot{\bf k})
-\frac{1}{3}(\mathbf{\sigma_1}\cdot\mathbf{\sigma_2)}
\big[{\bf k'}\delta({\bf r_1 - r_2}){\bf k}\big]\bigg\},
\label{tensor1}
\end{eqnarray}
where, as usual, ${\bf k}=({\bf \nabla}_1 - {\bf \nabla}_2)/2i$ acts on the right and ${\bf k'}=-({\bf \nabla}_1 - {\bf \nabla}_2)/2i$ on the left.
In the case of Skyrme forces the tensor interactions contribute to both binding energy and spin-orbit potential owing to their
dependence on the neutron and proton spin densities, $J_n$ and $J_p$, respectively \cite{Colo2007,Brink2018}. However, in the case of
the Gogny interaction the spin densities only appear in the spin-orbit energy density within the QLDFT formalism. Therefore the tensor term only 
modifies the spin-orbit part of the energy density. As we are using a short-range tensor interaction, the associate energy density   
will read \cite{Colo2007,Brink2018}
\begin{equation}
{\cal H}_T = \frac{1}{2}\alpha_T \bigg[ {\bf J}_n^2 + {\bf J}_p^2 \bigg] + \beta_T {\bf J}_n{\bf J}_p,
\label{tensor2}
\end{equation}
where the coefficients $\alpha_T$ and $\beta_T$ are related to the tensor strengths by
\begin{equation}
\alpha_T = \frac{5}{12} U \qquad \beta_T = \frac{5}{24} (T + U).
\label{tensor3}
\end{equation} 
If the tensor term is included, the form factor of the spin-orbit potential for each type of particles (given by the variations with respect to the spin 
densities  ${\bf J}_n$  and ${\bf J}_p$) is modified and reads 
\begin{equation}
{\bf W}_q  = \frac{W_o}{2} \bigg(2{\bf \nabla} \rho_q  + {\bf \nabla} \rho_{q'}\bigg)
+ \alpha_T{\bf J}_q + \beta_T{\bf J}_{q'} 
\label{tensor4}
\end{equation} 

\section*{Results and Discussion}
\subsection*{Impact of the incompressibility on the crossing of single particle levels in neutron rich Ni isotopes}
\begin{figure}[!tbp]
	\begin{minipage}[b]{0.45\textwidth}
		\includegraphics[width=1.2\textwidth]{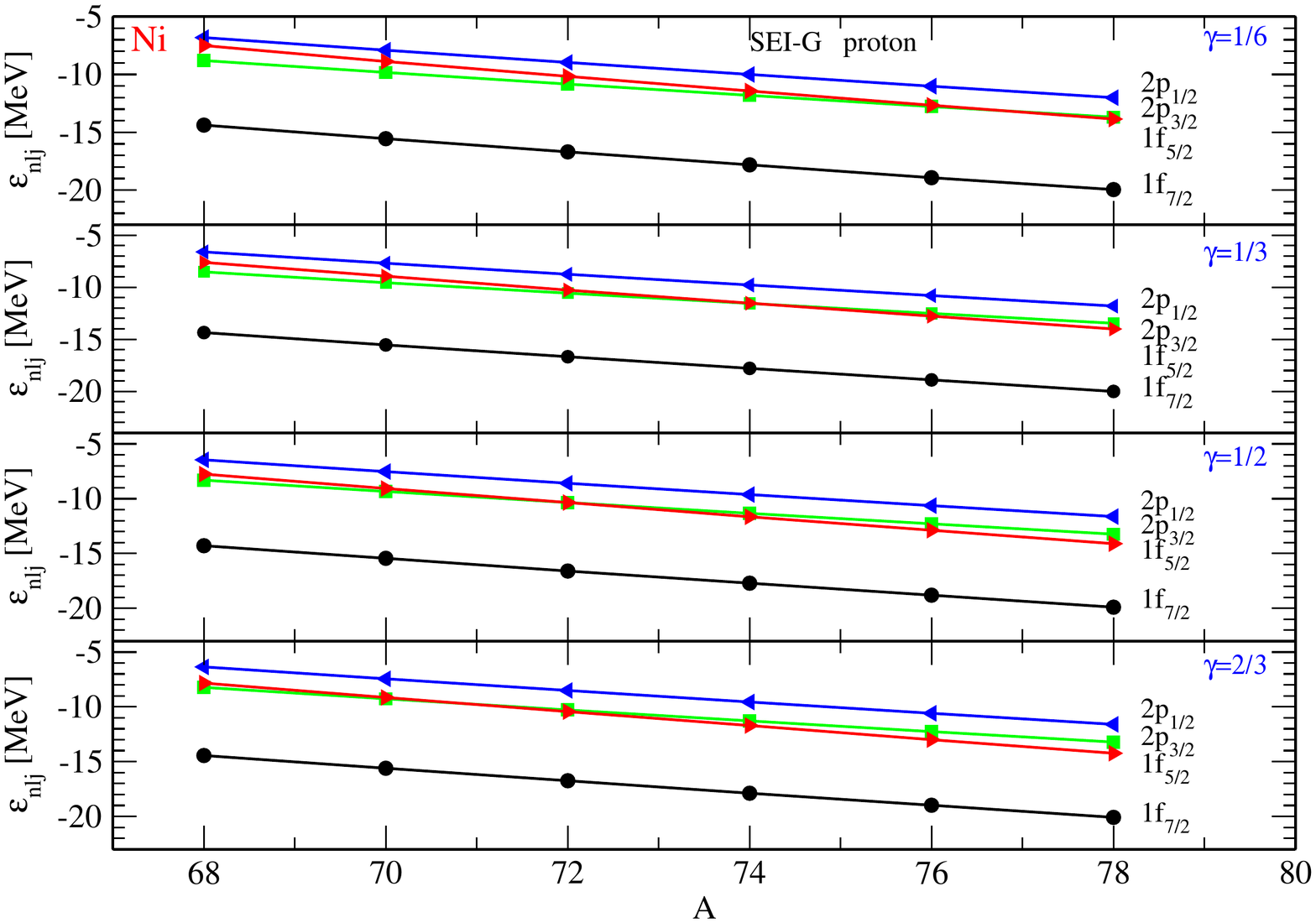}
		\caption{Proton single-particle levels around the Fermi level for
			Ni isotopes from A = 68 to A = 78 computed with the SEI interaction for the four EoS.}
		\label{Nisinglep}
	\end{minipage}
	\hfill 
	\begin{minipage}[b]{0.45\textwidth}
		\includegraphics[width=1.2\textwidth]{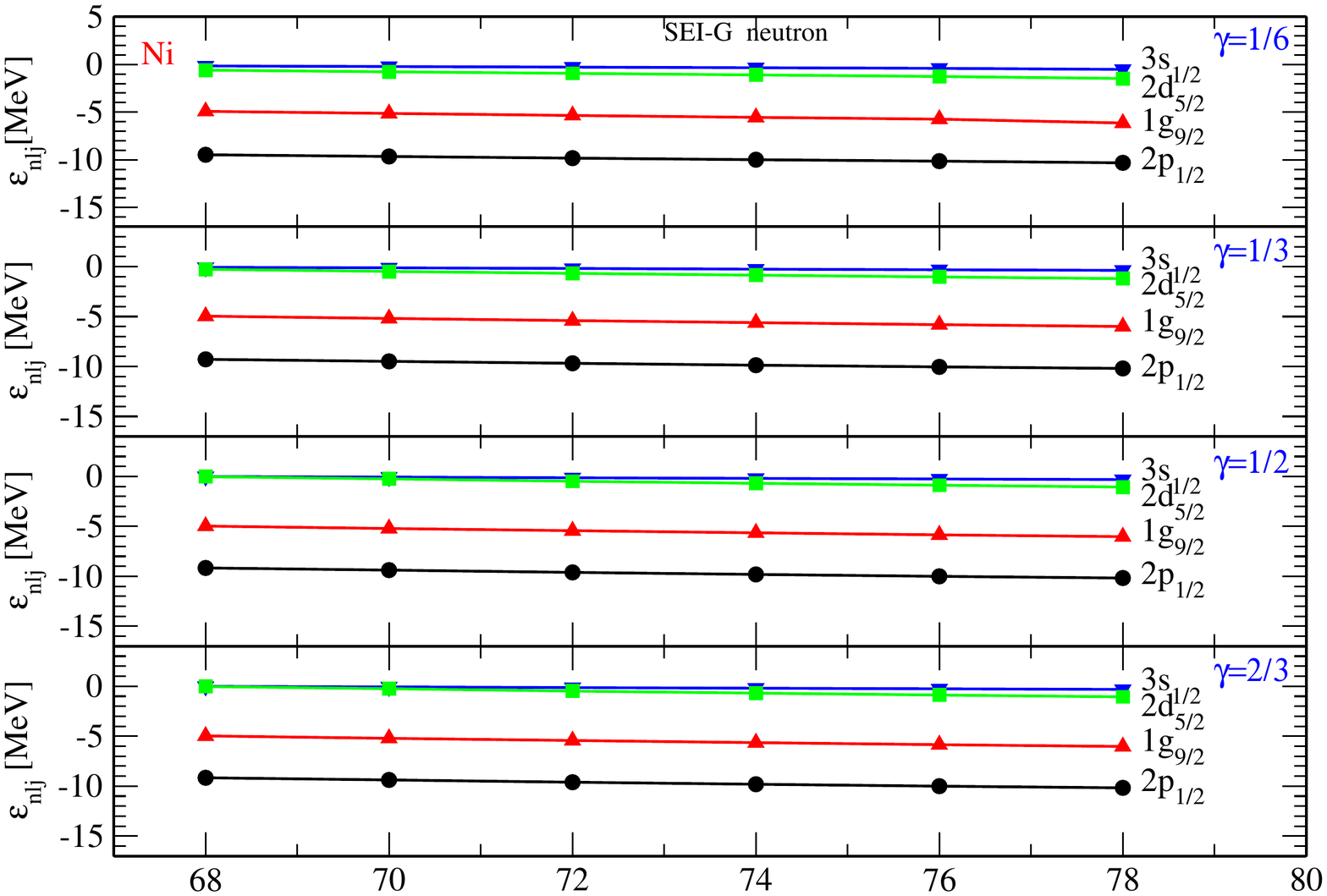}
		\caption{Neutron single-particle levels around the Fermi level for
			Ni isotopes from A = 68 to A = 78 computed with the SEI interaction for the four EoS.}
		\label{Nisinglen}
	\end{minipage}
\end{figure}

In this work we shall examine first the influence of NM incompressibility on the $1f_{5/2}$ and $2p_{3/2}$ s.p. levels crossing phenomenon 
predicted in $Ni$-isotopes. For this purpose we have used the 2015 parametrization of SEI, where all the parameters are the same 
as Table 2 
of Ref.\cite{Behera2015} except $x_{0}$ and $W_{0}$, which are fitted to reproduce the effective mass splitting in spin polarized PNM 
predicted by microscopic DBHF calculations and the  BE of $^{208}Pb$, respectively. The corresponding values for the different SEI EoS 
sets used in this work are reported in Table 4 of the same reference. 
The proton single particle (s.p.) energies of $1f_{7/2}, 1f_{5/2}, 2p_{3/2}$ and $2p_{1/2}$ in $Ni$-isotopes for $N$=40 to $N$=50 
have been 
calculated under the QLDFT formulation using the four EoSs corresponding to $\gamma=\frac{1}{6}, \frac{1}{3}, \frac{1}{2}$, and $\frac{2}{3}$ 
and the corresponding results are displayed in the Fig.\ref{Nisinglep}. The neutron s.p. energies for $2p_{1/2}, 
1g_{9/2}, 2d_{5/2}$ and $3s_{1/2}$, are 
shown in the Fig.\ref{Nisinglen}. Under the present formulation using SEI, the crossing of $1f_{5/2}$ and $2p_{3/2}$ s.p. proton levels in the 
isotopic chain of $Ni$ is found to be a function of NM incompressibility, as can be appreciated from the results displayed in Fig.\ref{Nisinglep}
for the four considered EoSs.  
As it is discussed in detail in Ref.\cite{Routray2021}, some effective forces of Skyrme and Gogny types can reproduce the crossing between the 
$1f_{5/2}$ and $2p_{3/2}$ proton s.p. level at mass number $A$=74, in agreement with the experimental data, only if an additional  tensor force 
is added to these interactions. The main effect of the tensor force in the $Ni$ isotopes analyzed in this work is the attraction (repulsion) between 
the neutron $1g_{9/2}$ s.p. level, whose occupancy grows when the mass number increases, and the  $1f_{5/2}$ ($2p_{3/2}$) s.p. proton level. Notice that the presence of a tensor term does not guarantee the crossing of these s.p. proton levels, as is the case of the 
Skyrme-Lyon force SLy5, as has been discussed recently in \cite{Routray2021}. This fact point out the relevance of the underlying mean-field
to describe the crossing of the aforementioned s.p. proton levels. For example in the case of SEI, the crossing between the $1f_{5/2}$ and the 
$2d_{3/2}$ s.p. proton levels is predicted at the right mass number if a EoS with a $\gamma$-value close to 1/2 ($K$=245 MeV) is used. SEI EoSs with 
smaller (larger) value of the incompressibility modulus $K$ move the crossing point towards higher (lower) mass numbers. For instance, the EoS 
$\gamma=\frac{1}{6}$ (K=207 MeV), predicts the crossing at $A$=78 while a SEI EoS with $\gamma$-value between 1/2 ($K$=245 MeV) and 2/3 ($K$=263 MeV) 
at $A$=72.
\begin{figure}
	\begin{center}
		\includegraphics[height=6cm,width=15cm]{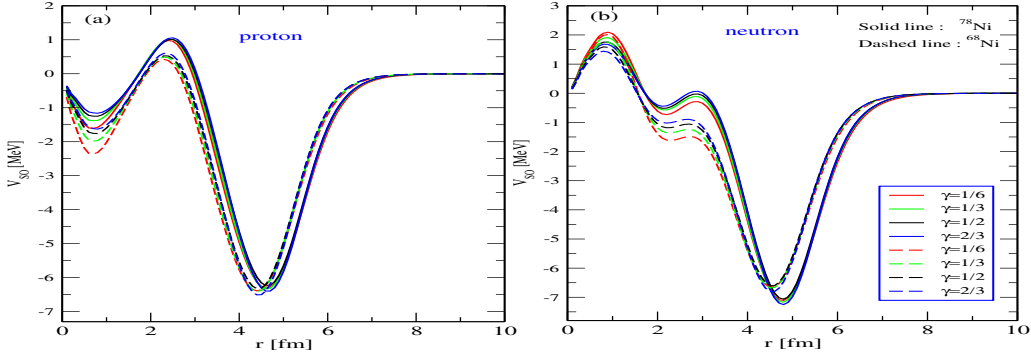}
		\caption{Proton and neutron SO-contributions in $^{68}Ni$ and $^{78}Ni$ as a function of distance r from the center of the nucleus for the four EoS.}
		\label{Ni_g12VSOpn}	
	\end{center}
\end{figure}
\begin{figure}
	\begin{center}
		\includegraphics[height=6cm,width=15cm]{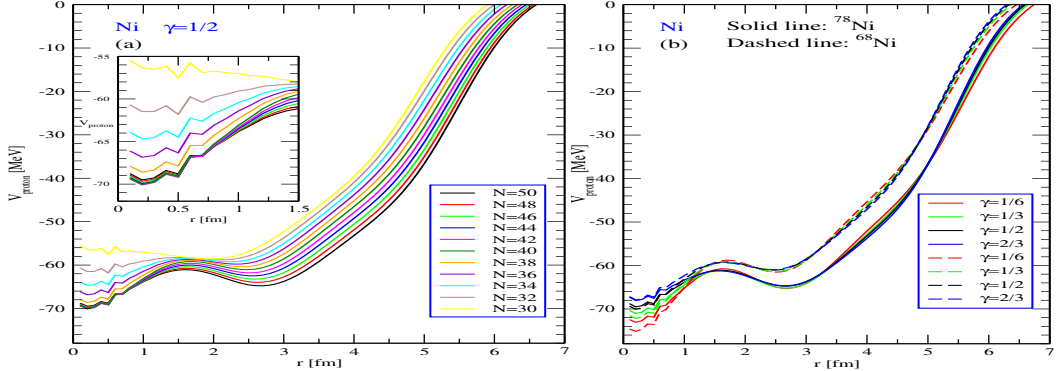}
		\caption{(a)Proton mean-field for Ni(N=30-50)isotopes predicted by the SEI for $\gamma=\frac{1}{2}$ under the QLDFT formulation.(b)Proton mean-field for $^{68}Ni$ and $^{78}Ni$ isotopes predicted by the SEI for $\gamma=\frac{1}{6}$, $\frac{1}{3}$, $\frac{1}{2}$ and $\frac{2}{3}$ under the QLDFT formulation.}
		\label{Ni_g16131223Vp}	
	\end{center}
\end{figure}

From Fig.\ref{Nisinglep} we can also see that the energy gap between the $1f_{5/2}$ and $1f_{7/2}$ proton-shells remains nearly 
stationary for all the four EoSs of SEI.
As one moves from $^{68}Ni$ to $^{78}Ni$, the proton energy gap, $(1f_{5/2}-1f_{7/2})$, 
decreases by 0.779 MeV, 0.744 MeV, 0.741MeV and 0.762 MeV for the EoSs $\gamma$=$\frac{1}{6}$, $\frac{1}{3}$, $\frac{1}{2}$, and $\frac{2}{3}$, 
respectively. Thus, for all the EoSs of SEI the magicity of Z=28 is preserved. The energy of the $1g_{9/2}$ and $2d_{5/2}$ s.p. neutron levels 
predicted by the different EoSs of SEI remain almost stationary as one moves from $^{68}Ni$ to $^{78}Ni$, as can be seen in 
Fig.\ref{Nisinglen}. This implies that the magicity of the neutron number N=50 is also preserved. 
To get more  insight about the evolution of shell structure predicted by SEI in Ni isotopes, we shall analyze the central and spin-orbit  
contributions to the neutron and proton mean-fields. 
In panels (a) and (b) of Fig.\ref{Ni_g12VSOpn} we display the SO contributions to p- and n- mean-fields as a function of the distance from the 
center for the nuclei $^{68}Ni$ and $^{78}Ni$ calculated with the four SEI EoSs used in this work.
For all these four EoSs, the results are similar in quality with closely lying values. 
This behaviour is different from the one obtained with the D1M Gogny interaction at QLDFT level. In that case, the SO-contribution to the 
p-mean-field along this $Ni$-isotopic chain shows a highly repulsive behaviour inside the nuclear volume as mass number increases, which is damped 
by the inclusion of the tensor part to the interaction that, on the other hand, has a negligibly small effect on the central potential. 
In panel (a) of Fig.\ref{Ni_g16131223Vp} we display the central part of the p-mean field for the $Ni$-isotopes from $N$=30 
to 50 as a function of the distance from the center $r$ computed with the EoS $\gamma=\frac{1}{2}$. We find a rather collectively compact behaviour 
of the p-mean-field potential curves for the isotopes corresponding to the filling of the $1g_{9/2}$ neutron shell. All the curves corresponding to 
$N$=41 to 50 starts from nearly identical value from well within the nuclear volume and varies smoothly as $r$ increases, predicting a relatively 
higher attractive potentials for higher mass isotopes. Such collective behaviour is not noticed for the filling of $1f_{5/2}$ or $2p_{3/2}$ 
neutron shells in the $Ni$-isotopes in the mass range $40\geqslant A \geqslant30$. We have also checked that the D1M Gogny force at QLDFT level
does not predict such a collective behaviour in the p-mean field potential curves for the $Ni$-isotopes when the neutron $1g_{9/2}$ is progressively
occupied. The collective behaviour of the p-mean field in the $Ni$-isotopes in range $N$=41 to 50 is qualitatively the same at the surface for the 
four EoSs of SEI corresponding to $\gamma$=$\frac{1}{6}$, $\frac{1}{3}$, $\frac{1}{2}$ and $\frac{2}{3}$, however, some small differences appear 
in the nuclear volume, which  is relatively more attractive for the EoS having lower value of incompressibility. This can be seen in panel (b) of 
Fig.\ref{Ni_g16131223Vp}, where the p-mean fields for the $^{68}$Ni and $^{78}$Ni are displayed for the four EoSs of SEI considered in this work.\\

\begin{figure}[h]
        \begin{center}
                \includegraphics[height=7cm,width=12cm]{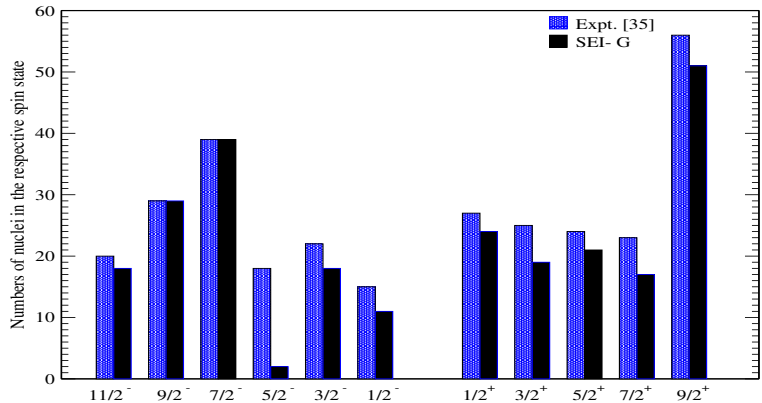}
                \caption{Comparison of experimental\cite{nndc} and SEI-G($\gamma=0.42$) spins of 298 odd-nuclei in different spin states.}
                \label{hist}
        \end{center}
\end{figure}

The experimental data on the s.p. energy levels in $Ni$-isotopes are scanty and the possible information is extracted from the studies in $Co$, $Zn$, 
and, $Cu$-isotopes. In recent $\gamma$-spectroscopic studies \cite{Olivier,Sahin} of $Cu$-isotopes, it has been found that the inversion of 
the ground state spin-parity from 3/2$^-$ to 5/2$^-$ occurs at $N$=46. This finding suggests the crossing of $2p_{3/2}$ and $1f_{5/2}$
proton s.p. levels in the underlying $Ni$-core. In order to study these $Cu$ isotopes, we show first that spherical odd nuclei can also be described 
fairly well using SEI at QLDFT level with the uniform blocking method of Ref.\cite{Perez2008}. The spin and parity of 298 spherical odd-nuclei have 
been computed with the EOS of SEI of $\gamma$=0.42 (see below), and the results are displayed in Figure \ref{hist}. More than 80\% of the experimental
 data of spin-parity of the ground-state of the odd-nuclei considered are predicted. This result is similar to the compilation of spins 
performed in Ref.\cite{bonneau2007} based on the results provided by several Skyrme forces and the FRDM of M\"oller.
To investigate the inversion of the spin-parity in $Cu$ isotopes, we have calculated the energy of the ground and several s.p. excited states of  
$Cu$-isotopes in the mass number region $A$=69-79 assuming spherical symmetry for all the considered nuclei because the  deformation 
in these 
neutron-rich $Cu$-isotopes is small \cite{Hilaire2007}. The ground state energies and spin-parity of the $Cu$-isotopes together with their first 
excited state energies are collected in Table \ref{Cu_BE} for the EoS $\gamma$=0.42.\\
\begin{table}
	\caption{Ground-state spin and energy of neutron-rich odd $Cu$ isotopes predicted by the SEI EoS ($\gamma$=0.42). The energy of the first excited state E* is also given along with the experimental results taken from Ref.\cite{Olivier}.}
	\begin{center}
		\begin{tabular}{cccccc}
			\hline
			
			\hline
			Nucleus&Spin-parity&SEI($\gamma=0.42$)&Expt.&SEI($\gamma=0.42$)&Expt.\\
		    &&Energy[MeV]&Energy[MeV]&E*[keV]&E*[keV]\\
			\hline
			$^{69}$Cu&3/2-&-599.40&-599.97&663&1215\\
			$^{71}$Cu&3/2-&-613.73&-613.09&449&537\\
			$^{73}$Cu&3/2-&-626.51&-625.51&156&263\\
			$^{75}$Cu&5/2-&-638.25&-637.13&103&62\\
			$^{77}$Cu&5/2-&-649.11&-647.42&292&295\\
			$^{79}$Cu&5/2-&-658.94&-656.65&620&660\\
			\hline
			\label{Cu_BE}	
		\end{tabular}	
	\end{center}
\end{table}
The inversion of the ground-state spin-parity from 3/2$^-$ to 5/2$^-$ in $Cu$ isotopes moves to smaller mass numbers when the incompressibility
modulus of the SEI EoS increases. To get the right inversion point, we have varied the $K$($\rho_{0}$) value of the EoS by varying the exponent 
$\gamma$, and it has been found that for $\gamma$=0.42 the inversion of the spin-parity of the ground state occurs at $N$=46, in agreement with
the experimental results \cite{Olivier,Sahin}, as it can be seen in Table \ref{Cu_BE}. This EoS also predicts reasonably well the energy of the 
first excited state, as can also be seen in the same Table \ref{Cu_BE}. 
The  calculated energies of the first excited states in other $Cu$-isotopes in this chain also compares well with the experimental results taken 
from Fig.2 of Ref.\cite{Olivier}. For example, the nucleus $^{75}Cu$ described by this EoS of SEI predicts the energy of excited state 3/2$^-$ 
at 103 KeV, whereas, the experimental value is 62 KeV. 
The analysis of the experimental data of $^{79}Cu$ also suggests another excited state 1/2$^-$ 1511 keV above its ground state, while the 
excitation energy of this state is 1957 keV according to large shell-model calculations \cite{Olivier} and it is predicted to be 
2254 keV by the SEI EoS ($\gamma$=0.42). This EoS of SEI with $\gamma$=0.42, which predicts the inversion of the spin-parity of the ground-state
from 3/2$^{-}$ to 5/2$^{-}$ for the nucleus $^{75}Cu$ and  gives a satisfactory description of excited states in $Cu$-isotopes, has NM 
incompressibility modulus of $K(\rho_{0})$=240 MeV. This value also conforms to the range for $K(\rho_{0})=240\pm20$ MeV extracted from the 
compressional mode of vibration in finite nuclei \cite{Shlomo2006} and the ranges obtained from allied studies of isoscalar giant monopole 
resonance (ISGMR) and heavy-ion collisions in finite nuclei \cite{Stone2014,Piekarewicz2009,Li2007,Wang2018}. 
\begin{table}[h]
	\caption{\small{Twelve numbers of interaction parameters for SEI-G($\gamma=0.42$) along with the nuclear matter saturation properties (such as saturation density $\rho_{0}$($fm^{-3}$), energy per nucleon $e(\rho_{0})$(MeV), incompressibility for symmetric nuclear matter K(MeV), effective mass m*/m, Symmetry energy $E_{s}$(MeV), Slope of symmetry energy L(MeV) and curvature of the symmetry energy $K_{sym}$(MeV)).}}
	\begin{center}
		
		\begin{tabular}{cccc}
			\hline\hline
			$\gamma$&b[$fm^{3}$]&$\alpha$[fm]&$\varepsilon_{ex}$[MeV]\\
			\hline
			0.42&0.5050&0.7591&-95.0536\\
			\hline
			$\varepsilon_{ex}^{l}$[MeV]&$\varepsilon_{0}$[MeV]&$\varepsilon_{0}^{l}$[MeV]&$\varepsilon_{\gamma}$[MeV]\\
			\hline
			-63.3691&-91.6562&-53.1272&90.0035\\
			\hline 
			$\varepsilon_{\gamma}^{l}$[MeV]&$t_{0}$[MeV$fm^{3}$]&$x_{0}$&$W_{0}$[MeV]\\
			\hline
			65.3966&341.2&1.7933&113.4\\
			\hline
			\multicolumn{4}{c}{\textbf{Nuclear matter saturation properties}}\\
			\hline
			$\rho_{0} [fm^{-3}]$&$e(\rho_{0})$[MeV]&K[MeV]&m*/m\\
			\hline
			0.1584&-16.0&240&0.711\\
			\hline
			$E_{s}$[MeV]&L[MeV]&$K_{sym}$[MeV]&\\
			\hline
			35.5&76.71&-155.0\\
			\hline\hline 
			\label{parameters}	
		\end{tabular}
	\end{center}
\end{table}
 The parameters of this SEI EoS of $\gamma$=0.42 alongwith the saturation properties are given in Table \ref{parameters}.
\begin{figure}[!tbp]
	\begin{minipage}[b]{0.45\textwidth}
		\includegraphics[width=1.2\textwidth]{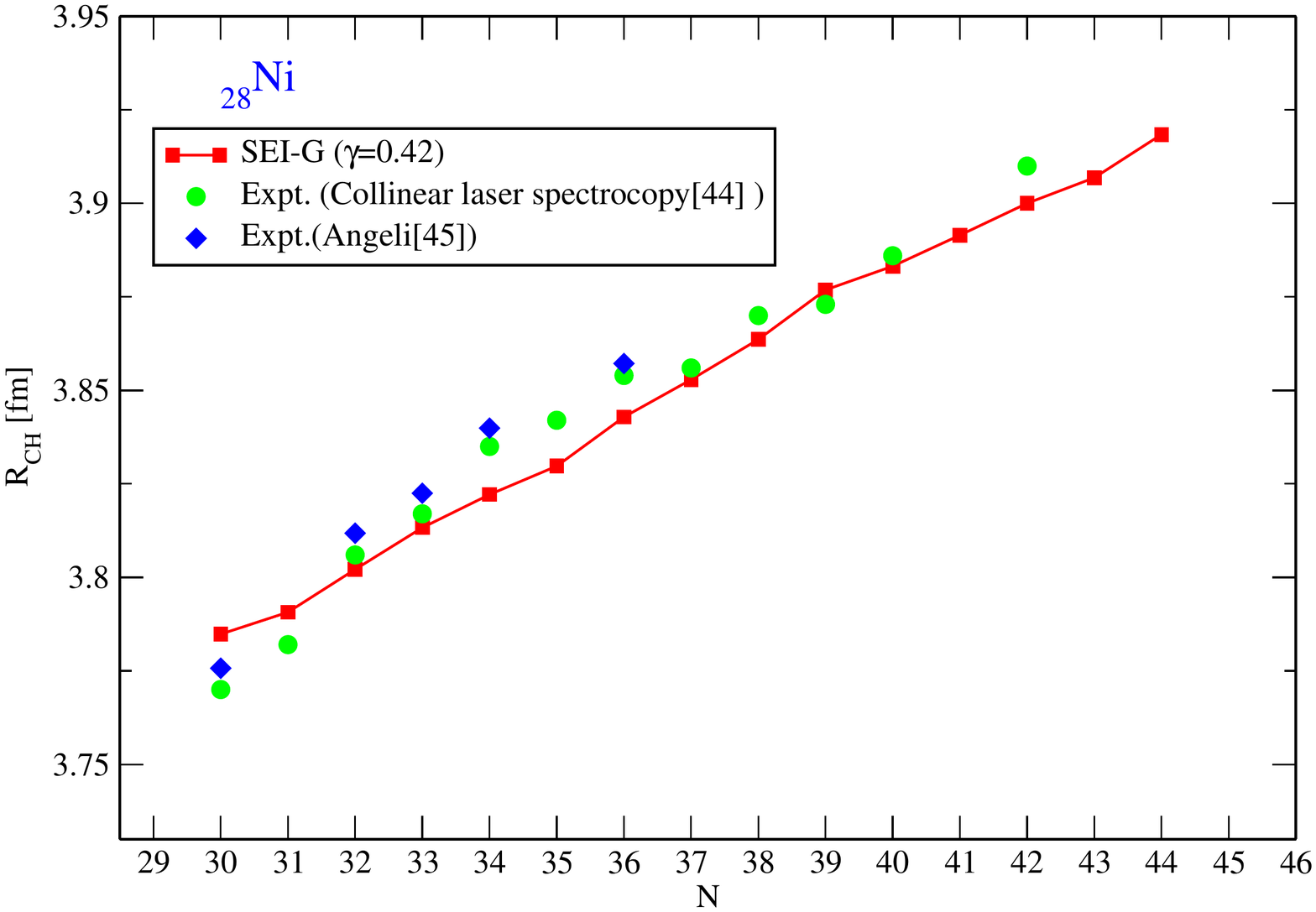}
		\caption{\tiny{Nuclear charge radii $R_{CH}$ using SEI-G ($\gamma=0.42$) compared with the experimental data \cite{Malbrunot-Ettenauer2022} and \cite{Angeli2013}).}}
		\label{Ni_R_c}
	\end{minipage}
	\hfill 
	\begin{minipage}[b]{0.45\textwidth}
		\includegraphics[width=1.25\textwidth]{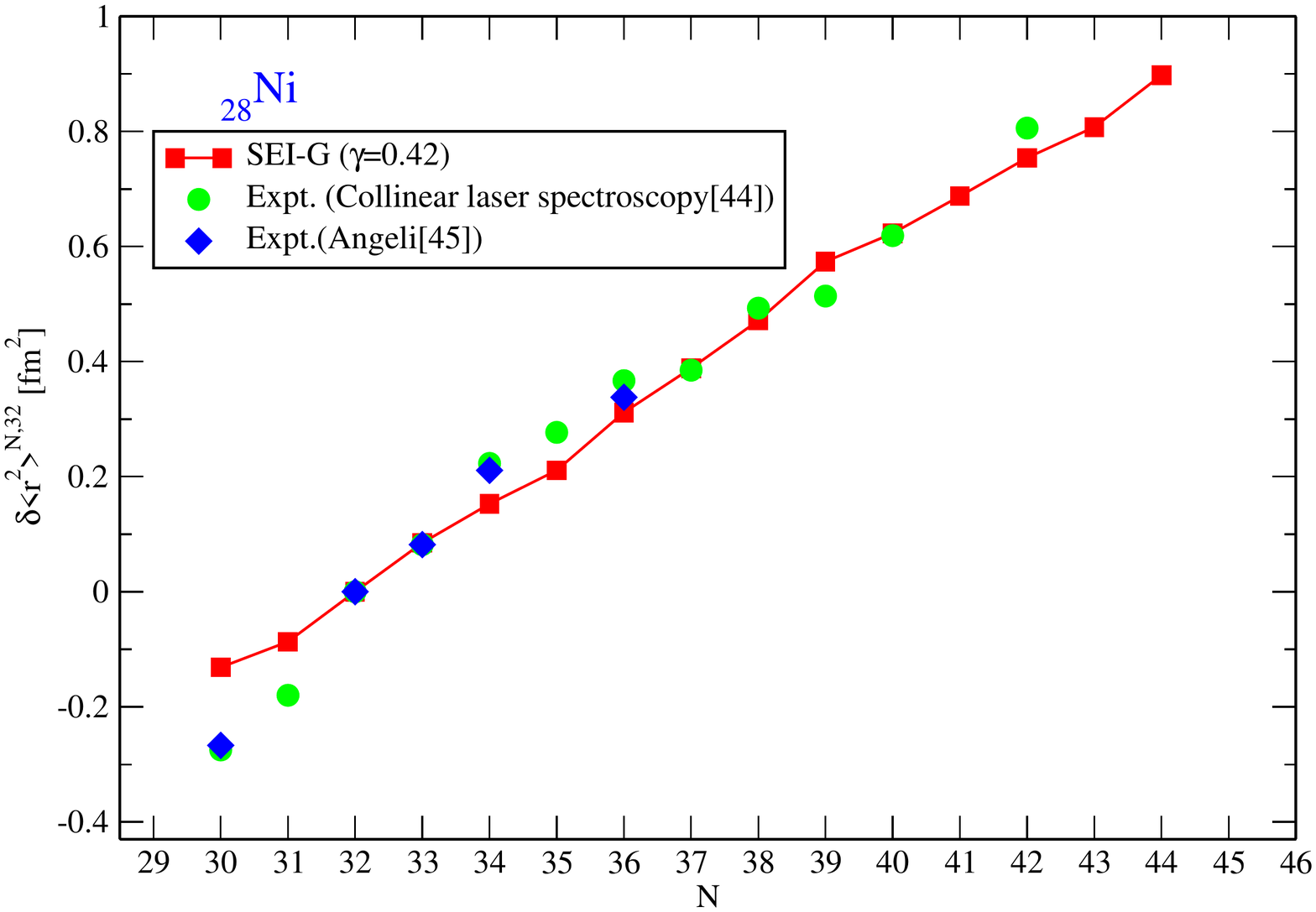}
		\caption{\tiny{Isotopic shift using SEI-G ($\gamma=0.42$) compared with the experimental data \cite{Malbrunot-Ettenauer2022} and \cite{Angeli2013}).}}
		\label{Ni_iso}
	\end{minipage}
\end{figure}
Very recently, measurements of charge radii $R_{CH}$ in $Z$=28 isotopes have been made \cite {Malbrunot-Ettenauer2022}, 
which allow to perform comparisons between experimental values and theoretical predictions of the charge radii in all light mass 
isotopic chains  
from $Z$=19 to $Z$=50. The charge radii $R_{CH}$ and the isotopic shifts $\delta{<r^2>}$ computed with the SEI EoS with $\gamma$=0.42 
of $Ni$-isotopes from $^{58}$Ni to $^{72}$Ni are displayed in the two panels of Fig.\ref{Ni_R_c} and Fig.\ref{Ni_iso} together with the 
experimental data \cite{Malbrunot-Ettenauer2022,Angeli2013}.
The SEI results are in good agreement with the experimental data extracted using the Collinear laser spectroscopy and showing a similar quality as predicted by the {\it {ab initio}} calculations using the  $NNLO_{sat}$ potential \cite{Malbrunot-Ettenauer2022}.\\
	\begin{figure}[h]
		\begin{center}					\includegraphics[height=7cm,width=12cm]{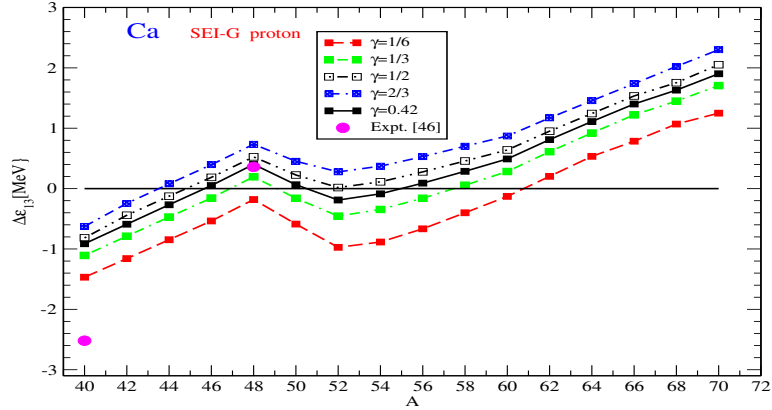}
			\caption{Energy differences of $2s_{1/2}$ and $1d_{3/2}$ proton levels (${\Delta\epsilon}_{13}$) of the Ca isotopes  predicted by the SEI for $\gamma$ = $\frac{1}{6}$ , $\frac{1}{3}$, $\frac{1}{2}$, $\frac{2}{3}$ and 0.42. The experimental data is also shown which are taken from the Ref.\cite{Grawe2007}.}
			\label{Ca40_48_p}	
		\end{center}
	\end{figure}
 The role of the incompressibility is also noticed in the study of ${\it{sd}}$ level splitting in $Ca$ isotopic chain using the SEI model. 
 Experimental studies \cite{Doll1976,Ogilvie1987} establish that the proton $2s_{1/2}$ and $1d_{3/2}$ s.p. levels inverts going 
from $^{40}$Ca to $^{48}$Ca. Theoretically this inversion has been analyzed using both non-relativistic and relativistic models 
\cite{Grasso2007,Wang2011,Nakada2013}. This inversion has been assigned to the action of the tensor force, 
and most of the effective interactions needs an additional tensor term to reproduce the level crossing effect. However, 
there are some interactions, for example, the SKI5 and SGII Skyrme forces, the D1M Gogny interaction, and the DDME1 and NL3 relativistic mean field
sets,  
which predict the inversion of the $2s_{1/2}$ and $1d_{3/2}$ proton s.p. levels in $^{48}$Ca without considering an extra tensor contribution. 
In Refs.\cite{Grasso2007,Wang2011} it is found that the 
evolution of the orbitals and hence the energy difference ${\Delta\epsilon}_{13}$=$2s_{1/2}$-$1d_{3/2}$ as neutron number increases 
in the $Ca$-isotopic chain follows a similar trend for different interactions,
irrespective of the inversion is predicted at $^{48}$Ca or not. The energy difference ${\Delta\epsilon}_{13}$ in 
$Ca$ isotopes is shown as a function of $N$ in Fig.\ref{Ca40_48_p} for the four sets of SEI, $\gamma$=$\frac{1}{6},\frac{1}{3},\frac{1}{2}$,and 
$\frac{2}{3}$ together with the results for the EOS $\gamma$=0.42. The same behaviour in the evolution of ${\Delta\epsilon}_{13}$ is 
observed in all the five sets of SEI EoSs. The gap ${\Delta\epsilon}_{13}$ increases from $^{40}$Ca to $^{48}$Ca, and then decreases 
to a minimum at $^{52}$Ca and increases thereafter. This is a global trend observed in all the non-relativistic and relativistic 
interactions, as it is discussed in Refs.\cite{Grasso2007,Wang2011}. For the SEI EoS $\gamma$=1/6, the inversion between the $1d_{3/2}$ and $2s_{1/2}$ proton levels in $^{48}$Ca, understood as a change of sign of 
${\Delta\epsilon}_{13}$, i.e. with the $1d_{3/2}$ proton level going below of the $2s_{1/2}$ one, does not occur predicting 
${\Delta\epsilon}_{13}$=-0.18 MeV, as it happens in case of SLy5+T interaction in Ref.\cite{Wang2011}. For remaining four EoSs, $\gamma$=$\frac{1}{3}$, 0.42,
$\frac{1}{2}$, and $\frac{2}{3}$, the  
inversion is predicted in $^{48}$Ca, 
with ${\Delta\epsilon}_{13}$ value ranging 
between 195 keV to 734 keV. For the EoS $\gamma$=0.42, ${\Delta\epsilon}_{13}$=401 keV as compared to the experimental value of about 360 keV 
\cite{Grawe2007,Burrows1995}. On going from $^{48}$Ca to $^{52}$Ca, again a new inversion between the position of the $1d_{3/2}$ and 
$2s_{1/2}$ proton levels occurs 
in case of the two EoSs $\gamma$=1/3 and 0.42, whereas for the EoSs $\gamma\geqslant 1/2$, although the energy gap ${\Delta\epsilon}_{13}$ decreases,
 the inversion does not take place. The occurrence of the inversion 
moving from $^{48}$Ca to $^{52}$Ca, predicted by the EoSs $\gamma$=1/3 and 0.42, is supported with the experimental indications 
related to first-forbidden $\beta$-decay measurements \cite{Baumann1998}:  the low-energy levels in $^{50}$K are dominated by (${\pi}d3/2$)$^{-1}$  (${\nu}d3/2$)$^{1}$ configuration which is an indication that the inversion is not present at 
$N$ = 31 in the $K$ chain and, thus, at $N$ = 32 in the $Ca$ chain. 
Beyond $^{52}$Ca, ${\Delta\epsilon}_{13}$ increases again for all the five parameterizations of SEI. This generates 
another inversion with $\gamma$=1/3 and 0.42 EoSs starting at $^{56}$Ca. In this context it is to be mentioned that in the Ref.\cite{Nakada2013} 
it has been claimed that this inversion should occur in heavier $Ca$ isotopes close to the two neutron drift limit, which are the 
predictions of the semirealistic M3Y-P5$^{'}$and M3Y-P7 models based on a realistic tensor force able to
reproduce the ${\Delta\epsilon}_{13}$ in $^{40-48}$Ca without adjustment.
\subsection*{SEI plus tensor force}
In the previous subsection we have discussed the crossing of $2p_{3/2}$ and $1f_{5/2}$ s.p. levels in $Ni$-isotopes and the inversion 
of the ground-state spin-parity in $Cu$-isotopes predicted by SEI without inclusion of a tensor term in the interaction. In this subsection we want to 
examine other findings those require to consider explicitly a tensor contribution to the SEI interaction eq.(\ref{1}). We will analyze first the 
the energy differences between $1h_{11/2}-1g_{7/2}$ proton s.p.levels in the $Z$=50 and $Z=51$ isotopic chains and between the $1i_{13/2}-1h_{9/2}$ 
s.p. neutron levels gap in the $N$=82 isotonic chain compared to the available data \cite{Schiffer2004}. Next, we discuss the reduction 
of the spin-orbit splittings of the s.p. $f$ and $p$ levels at the $N$=28 closure. To this end, we compare our predictions to the
experimental investigation of the $^{46}$Ar(d,p)$^{47}$Ar reaction in inverse kinematics\cite{Gaudefroy2006}. 

\subsubsection*{Proton gaps in the $Z$=50 and $Z=51$ isotopic chains and neutron gaps in the $N$=82 isotonic chain:}
The energy gaps between the $1h_{11/2}$ and $1g_{7/2}$ proton s.p. levels and between the $1i_{13/2}-1h_{9/2}$ s.p. neutron levels, 
above the $Z$=50 and $N$=82 closed shells, respectively, are investigated through QLDFT calculations using SEI in the $Sn$ and $Sb$ isotopic chains and
in the $N$=82 isotonic chain. The predictions of the SEI model alone in these scenarios are  unable to reproduce the behaviour and 
the results extracted by Schiffer et al from the analysis of different experimental data that are reported in Ref.~\cite{Schiffer2004}.
In order to get a better agreement with the results reported in \cite{Schiffer2004}, we add to SEI a short-range tensor term. As we 
have shown in
Section 2, the energy density associated to the tensor force depends on the neutron and proton spin densities (see Eq.(\ref{tensor2})), 
and its main effect is to modify the spin-orbit potential Eq.(\ref{tensor4}), which in turn modifies the relative position of the 
neutron and proton s.p. energy levels. The parameters $T$ and $U$ of the tensor force are chosen to describe the energy gaps given in 
Ref.~\cite{Schiffer2004} 
under the constraint that the crossing of $2p_{3/2}$ and $1f_{5/2}$ s.p. levels in $Ni$-isotopes at neutron number $N$=46 remains unchanged. 
For each pair of $T$ and $U$ values, the spin-orbit strength $W_o$ is readjusted to reproduce the experimental BE of $^{208}Pb$. 
Following this protocol we have found that for $T$=800 MeV, $U$=-140 MeV and a spin-orbit strength $W_0$=122 MeV, the $f$ and $p$ level crossing 
at $N$=46 in $Cu$- and $Ni$-isotopes remain unchanged. For higher (lower) values for $T$ and $|U|$, the crossing shifts to higher (lower) value of 
$N$ in case of $Ni$. The negative value for $U$ is supported by the conclusion of Brown et al \cite{Brown2006}, who found that short-range tensor forces with $\alpha_T$ $<$0 reproduce better the experimental data in $^{132}Sn$ and $^{114}Sn$.
Our procedure of determining the tensor parameters $T$ and $U$ is different from the strategy used in \cite{Stancu1977}, where the parameters of the tensor force are fitted to reproduce the spectra of $^{48}$Ca and  $^{56}$Ni. 
We have checked that the the spin-orbit splittings of these two nuclei computed within our 
approach give quite similar results to the ones displayed in Figs.1 and 2 of Ref.~\cite{Stancu1977}.\\
The tensor force provides an additional attraction between neutron and proton particle or hole 
states with spins $j_{>}=l+1/2$ and $j'_{<}=l'-1/2$ (or with $j_{<}=l-1/2$ and $j'_{>}=l'+1/2$) and repulsion with spins $j_{>}=l+1/2$ and 
$j'_{>}=l'+1/2$ (or with $j_{<}=l-1/2$ and $j'_{<}=l'-1/2$).  
The mechanism of the tensor force, which produces the apparent attraction and repulsion between the s.p. states, is discussed in 
the work in Ref.\cite{Colo2007}, also applies to the theoretical results of the present work.
The $\alpha_T$-part of the tensor contribution to the spin-orbit form factor Eq.(\ref{tensor4}), has a constant effect without any 
isospin dependence. The isospin dependence enters through the $\beta_T$-part of the tensor term. 
As the triple-odd tensor strength $U$ is negative,
the $\alpha_T$ parameter is also negative. Therefore, the contribution from this term to the spin-orbit form factor is negative (positive) 
depending on whether $\textbf{J}_q$ corresponds to a $j_{>}$ ($j_{<}$) state, as can be seen from Eq.(\ref{Eq_J}).
The $\beta_T$-term in Eq.(\ref{tensor4}) has a positive (negative) contribution to the spin-orbit form-factor if $\bf {J}_{q'}$ is 
$j_{>}$ ($j_{<}$), as far as $\beta_T$ is positive for our case. 
When both the contributions from $\alpha_T$- and $\beta_T$-terms are negative, the 
spin-orbit contribution, which itself is negative, increases, thereby increasing the splitting of the s.p. states. Thus the energy gap between 
the $j_{>}=l+1/2$ and $j'_{<}=l'-1/2$ of the two different s.p. states decreases. 
These tensor interactions are stronger between states with similar radial wavefunctions, i.e. with the same principal quantum number and the same orbital angular momentum because in this case there is a large overlap along the radial 
directions \cite{Otsuka2005}. \\
	\begin{figure}[h]
		\begin{center}
			\includegraphics[height=7cm,width=13cm]{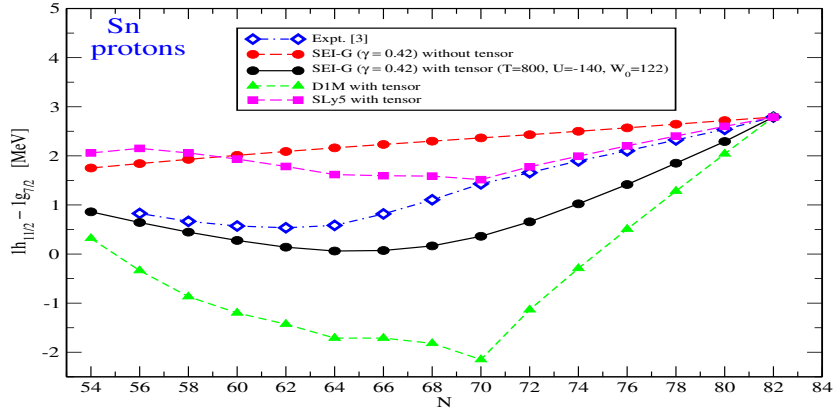}
			\caption{Energy differences between $1h_{11/2}$ and $1g_{7/2}$ proton s.p.levels in $Sn$-isotopes for SEI with and without Tensor. The experimental data are taken from Ref.\cite{Schiffer2004}. Theoretical results have been shifted so that splittings coincide with $^{132}$Sn. The corresponding results for D1M and SLy5 interaction sets with tensor are also given for comparison.}
			\label{Sn_h112_g72}
		\end{center}
	\end{figure}
	\begin{figure}[!tbp]
		\begin{minipage}[b]{0.5\textwidth}
			\includegraphics[width=1.1\textwidth]{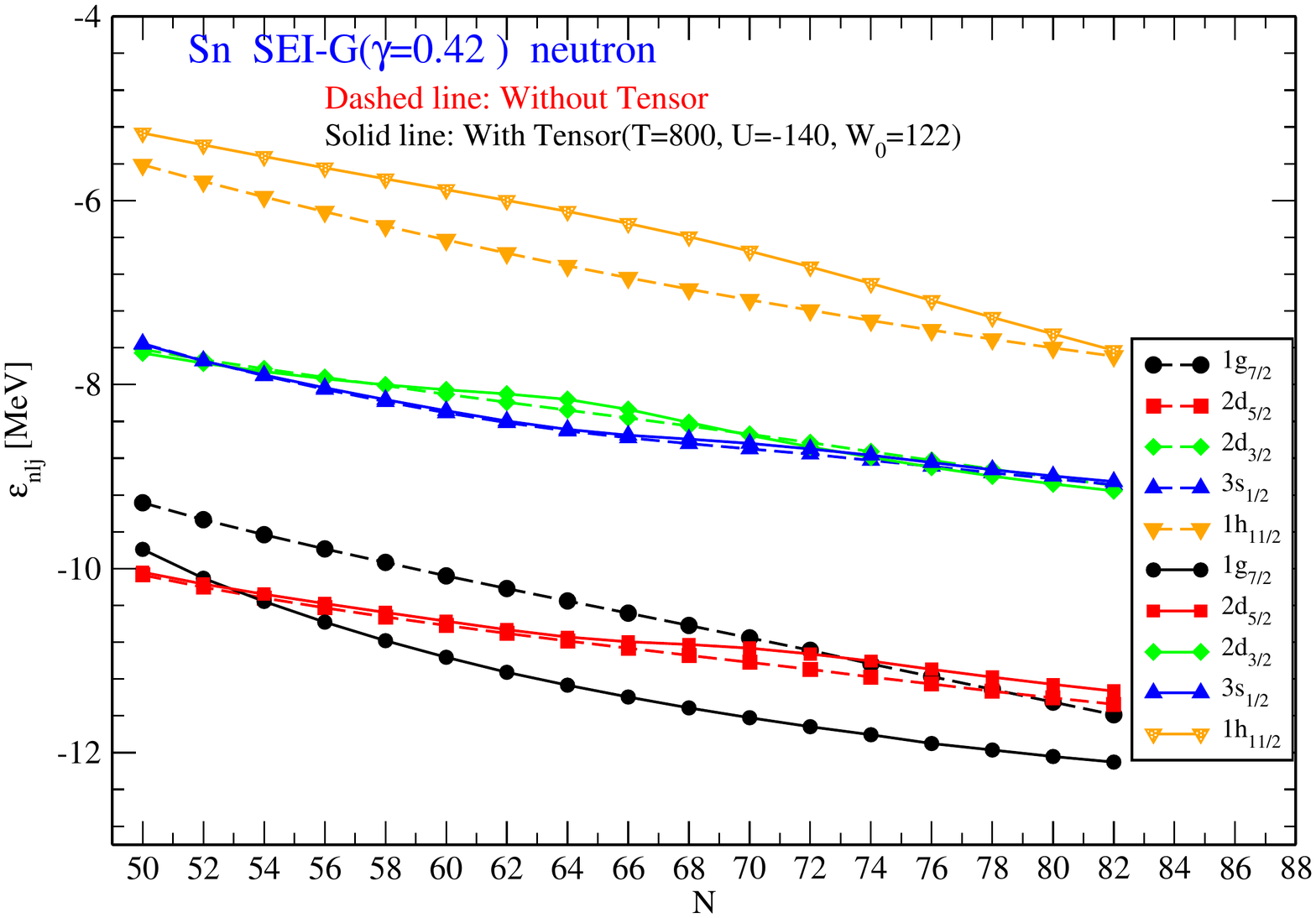}
			\caption{Neutron levels of Sn isotopes in the $N$=50 to $N$=82 major shell. Solid (dashed) lines correspond 
				to the s.p. energies computed with (without) tensor force.}
			\label{Sn-levels}
		\end{minipage}
		\hfill
		\begin{minipage}[b]{0.5\textwidth}
			\includegraphics[width=1.1\textwidth]{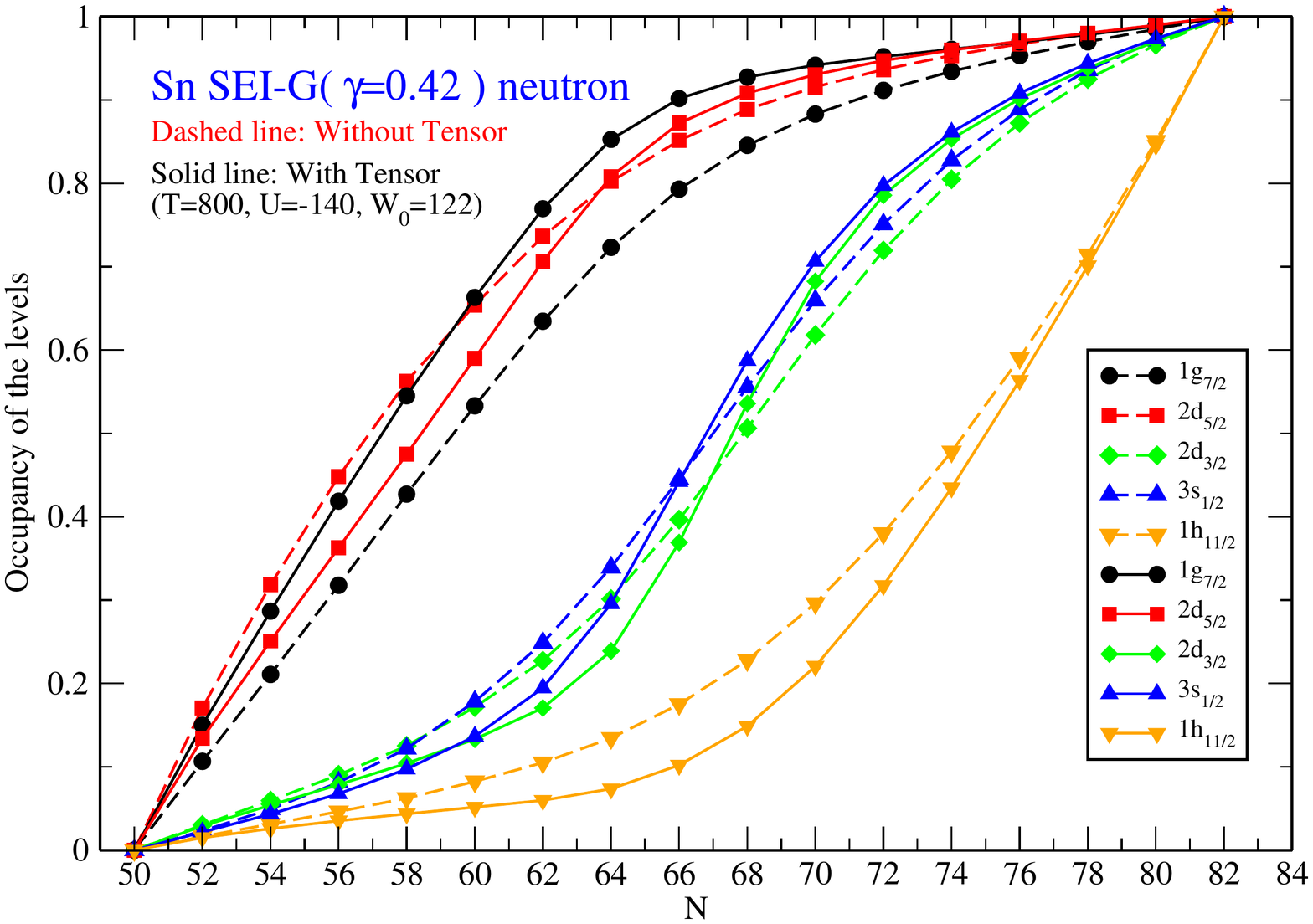}
			\caption{Occupation probability of the neutron levels of the Sn isotopes in the $N$=50 to $N$=82 major shell. Solid 
				(dashed) lines correspond to the occupations computed with (without) tensor force.} 
			\label{Sn-occupation}
		\end{minipage}
	\end{figure}
The tensor effects on the gap $1h_{11/2}$-$1g_{7/2}$ of the unoccupied proton states along the Sn-isotopic chain, shown in Fig.\ref{Sn_h112_g72}, 
strongly depend on the position and occupancy of the neutron s.p. levels, which in the case of SEI are displayed in Figs. \ref{Sn-levels} and \ref{Sn-occupation}, respectively. From Fig.~\ref{Sn-levels}, we see that the impact of the tensor force is more important on states of large orbital angular momentum, such as 
1$g_{7/2}$ or 1$h_{1172}$, whose s.p. energies are clearly shifted with respect to the values computed without the tensor interaction.
Above $N$=50, the neutron levels $1g_{7/2}$ and the $2d_{5/2}$ predicted by SEI lie very close to each other. The same 
happens with the $3s_{1/2}$ and $2d_{5/2}$ levels, while $1h_{11/2}$ remains isolate at higher s.p. energy. As can be seen from
Fig.~\ref{Sn-occupation}, from $A$=100 to $A$=114, neutrons in the Sn isotopic chain mainly populate the $1g_{7/2}$ and the $2d_{5/2}$ levels almost
 with the same occupation probability, which reaches at 80\% at $A$=114. Above this mass number the occupancy of the $3s_{1/2}$ and $2d_{3/2}$ levels 
increase remarkably until about 60\% in competition with the filling of the $1h_{11/2}$, which has a small occupation up to $A$=120, but from this 
mass number onward increases till saturate at $A$=132. 
The gap between the unoccupied $1h_{11/2}$ and $1g_{7/2}$ proton levels along the Sn-isotopic chain computed with and without the tensor 
interaction are shown in Fig.\ref{Sn_h112_g72} where we have shifted the theoretical results to coincide with the experimental value at $^{132}$Sn. This gives a better insight into the evolution of the orbitals under the influence of the tensor force along the $Sn$-isotopic chain \cite{Shen19}. We have also shown the corresponding results of the D1M and SLy5 interaction sets with tensor. For SEI the shift in the cases of with and without tensor part are $\delta$=2.99 MeV and 1.85 MeV, respectively. In the same figure we also show the experimental data extracted from ($\alpha,t$) reaction on tin 
isotopes displayed in the upper panel of Fig.3 of Ref.~\cite{Schiffer2004}. Due to the tensor force, the filling of the $1g_{7/2}$ neutron state 
enhances the spin-orbit splittings of the $h$ and $g$ s.p. proton states, thereby decreasing the gap between the 
$1h_{11/2}$ and the $1g_{7/2}$ proton levels. 
The effect of $2d_{5/2}$ neutron level on these two proton levels is just the opposite. As the occupancy of the $1g_{7/2}$ and $2d_{5/2}$ neutron levels is quite similar, there is a partial cancellation between the tensor effects of these two neutron 
levels and the $1h_{11/2}$ and $1g_{7/2}$ proton levels. Due to the large overlap between the s.p wavefunction of the $1g_{7/2}$ neutron 
state and the $1h_{11/2}$ or $1g_{7/2}$ proton states as compared with the overlap in the case of the $2d_{5/2}$ neutron state, the SEI calculation 
including tensor force predicts that the $1h_{11/2}$-$1g_{7/2}$ proton gap decreases when $A$ increases from 100 to 114, which is in agreement with 
the experimental data \cite{Schiffer2004}. When the mass number of the isotope increases above 
$A$=114, the occupancy of the $2d_{3/2}$ and $3s_{1/2}$ neutron levels becomes progressively important reaching about a 60\% at $A$=120. The tensor 
interaction  between the $2d_{3/2}$ neutron state and the $1h_{11/2}$ and $1g_{7/2}$ proton states should reduce the proton gap, however, owing to the $1h_{11/2}$ neutron level, which attracts the $1g_{7/2}$ and repels the $1h_{11/2}$ proton levels, the combined 
effect produces an increasing of the $1h_{11/2}$-$1g_{7/2}$ proton gap. 
In spite of the small occupancy of the $1h_{11/2}$ neutron 
level in the range between $A$=114 and $A$=120 (see Fig.~\ref{Sn-occupation}), the tensor interaction of this state with the $1h_{11/2}$ and 
$1g_{7/2}$ proton states is strong enough, due to its principal quantum number and large orbital angular momentum, to reverse the reduction of the 
gap due to the $2d_{3/2}$ neutron state. Finally above $A$=120, the occupancy of the $1h_{11/2}$ grows increasing the gap between the $1h_{11/2}$ and
 $1g_{7/2}$ proton states in agreement with the experimental trend \cite{Schiffer2004}. As it can be seen from Fig. \ref{Sn_h112_g72}, without tensor 
force, the SEI model predicts an almost linear smooth growing tendency with the neutron number in disagreement with the 
experimental results reported in \cite{Schiffer2004}, whose trends can be reproduced, at least qualitative, by adding the short-range 
tensor interaction.\\
 We can also analyzed the $1h_{11/2}$-$1g_{7/2}$ proton gap in the Sb isotopic chain, which behaves in a similar way as in the Sn chain discussed 
just before. The $1h_{11/2}$-$1g_{7/2}$ proton gap decreases when the neutron number moves from 50 to 68, owing the progressive occupation of the 
$1g_{7/2}$, $2d_{5/2}$ and $1s_{1/2}$ neutron levels and increases again above $N$=70 when the occupation of the $1h_{11/2}$ neutron level grows 
producing a gap of $\approx$6 MeV at $N$=82. 
Effects due to the tensor force can be seen in the evolution of the relative separation of the unoccupied $1i_{13/2}$-$1h_{9/2}$ neutron
levels in the isotones of $N$=82 in Fig.\ref{N82_1i132_1h92}. Like the $Sn$ isotope case we have shifted the theoretical results to coincide with the experimental value at $^{132}$Sn for a better display of the influence of the tensor force along the $N$=82 isotonic chain. We have also shown the corresponding results of the D1M and SLy5 interaction sets with tensor. For SEI the shift in the cases of with and without tensor part are $\delta$=1.164 MeV and 0.904 MeV, respectively. 
For this isotonic chain the evolution of the proton s.p. levels in the $Z$=50 to $Z$=72 
major shell and their corresponding occupancies as a function of the atomic number are displayed in Figs.~\ref{N82_p_levels} and \ref{N82_p_occu},
respectively. As proton number increases from $Z$=50 to $Z$=58, only the $1g_{7/2}$ proton level fills up until practically saturate at $Z$=58. Due 
to the tensor force, this proton level pulls the $1i_{13/2}$ and push the $1h_{9/2}$ neutron levels decreasing the gap between them. From $Z$=58 to 
$Z$=64 again only the $2d_{5/2}$ proton level fills up appreciably (see again Fig.~\ref{N82_p_occu}), but in this case the tensor force acts in the 
opposite way, i.e., push up the $1i_{13/2}$ and pull down the $1h_{9/2}$ neutron levels, increasing the $1i_{13/2}-1h_{9/2}$ neutron gap. The kink
at $Z$=62 is due to the increasing of the occupancy of $2d_{3/2}$ proton level, which compensates the enlarging of the gap due to 
the $2d_{5/2}$ level. Beyond $Z$=64 and upto $Z$=70, the situation in the case of SEI with EOS $\gamma$=0.42 is more complicated because the 
$2d_{3/2}$, $3s_{1/2}$ and $1h_{11/2}$ proton levels are almost degenerated and they populate simultaneously, as it can be seen in 
Fig.\ref{N82_p_occu}. In this scenario, the reduction of the $1i_{13/2}-1h_{9/2}$ neutron gap due to the $2d_{3/2}$ proton level is 
compensated by the increasing effect of the $1h_{11/2}$ proton level, whose wave-function has larger overlap with the wave-functions of the neutron 
levels because of the same principal quantum number and similar value of the orbital angular momentum.  
	 \begin{figure}[h]
	 	\begin{center}
	 		\includegraphics[height=7cm,width=12cm]{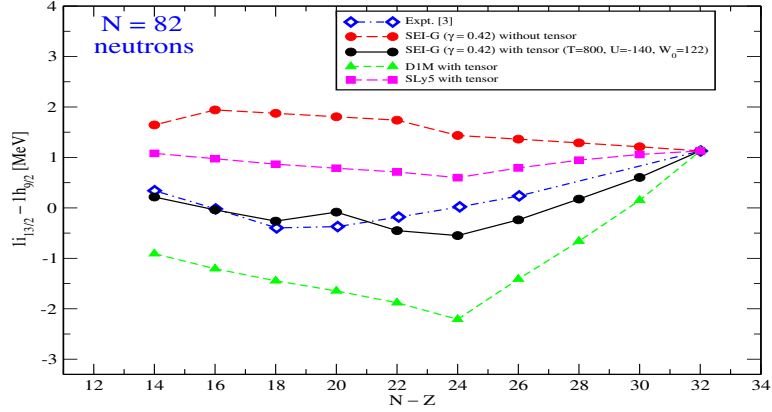}
	 		\caption{Energy differences between $1i_{13/2}$ and $1h_{9/2}$ neutron s.p. levels in $N=82$ computed for SEI with and without tensor interaction. The experimental data are taken from Ref.\cite{Schiffer2004}. Theoretical results have been shifted so that splittings coincide with $^{132}$Sn. The corresponding results for D1M and SLy5 interaction sets with tensor are also given for comparison.}
	 		\label{N82_1i132_1h92}
	 	\end{center}
	 \end{figure}
        \begin{figure}[!tbp]
                \begin{minipage}[b]{0.5\textwidth}
                        \includegraphics[width=1.1\textwidth]{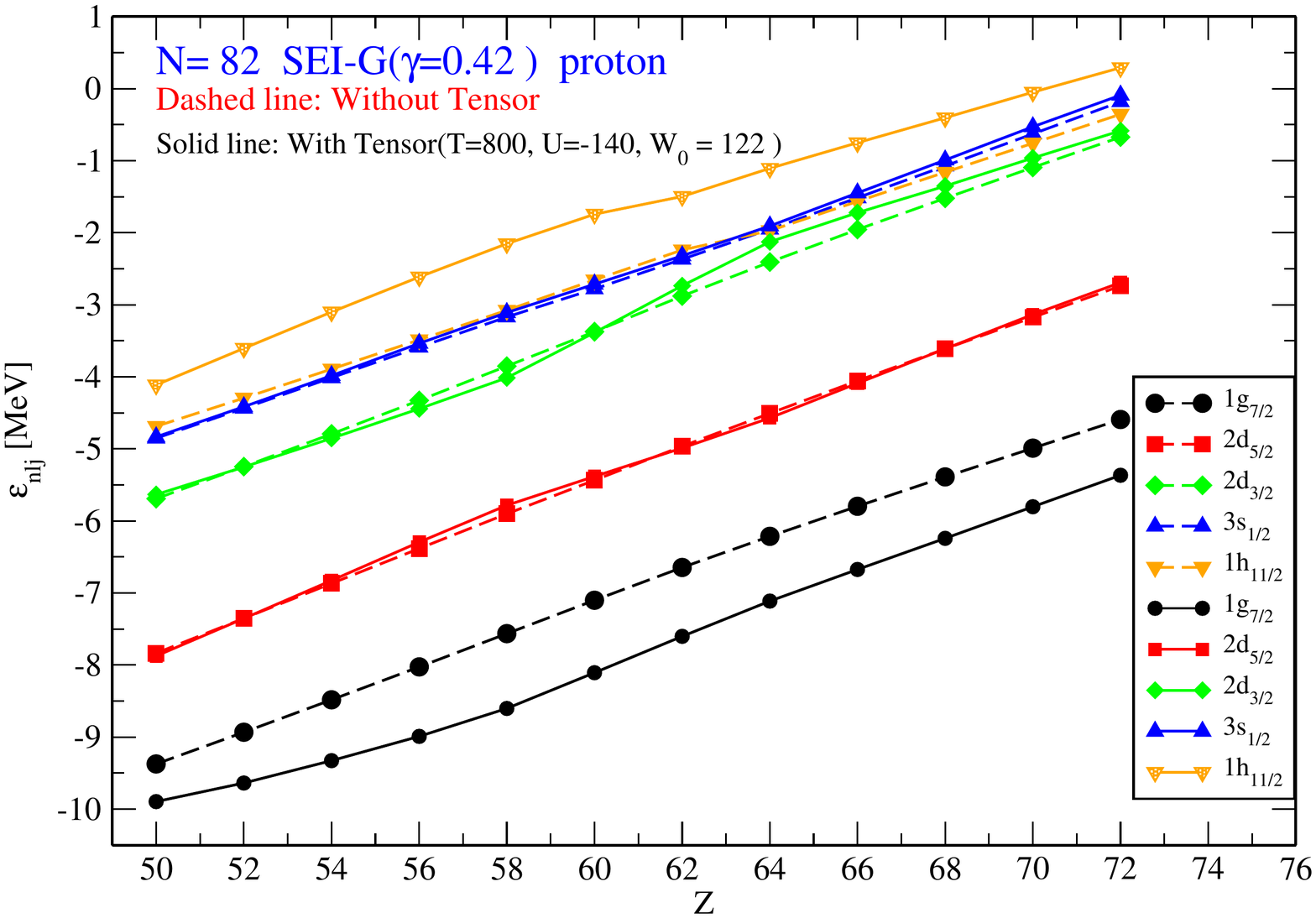}
                        \caption{Proton levels of the $N$=82 isotones in the $Z$=50-$Z$=72 major shell. Solid (dashed) lines correspond
                        to the s.p. energies computed with (without) tensor force.}
                        \label{N82_p_levels}
                \end{minipage}
                \hfill
                \begin{minipage}[b]{0.5\textwidth}
                        \includegraphics[width=1.1\textwidth]{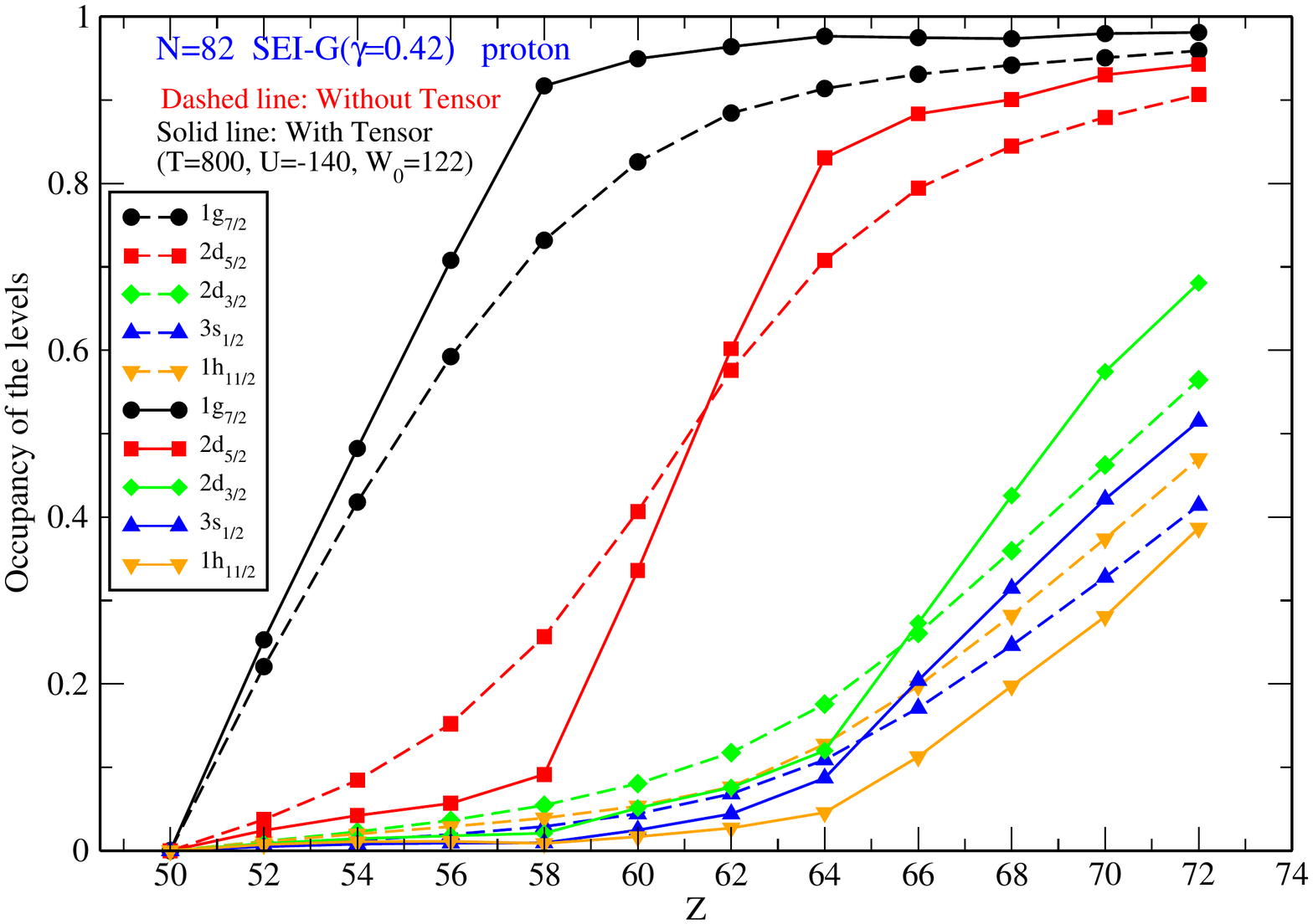}
                        \caption{Occupation probability of the proton levels of the $N$=82 isotones in the $Z$=50-$Z$=72 major shell. Solid
                       (dashed) lines correspond to the occupations computed with (without) tensor force.}
                        \label{N82_p_occu}
                \end{minipage}
        \end{figure}
\begin{figure}[h]
	\begin{center}
		\includegraphics[height=7cm,width=12cm]{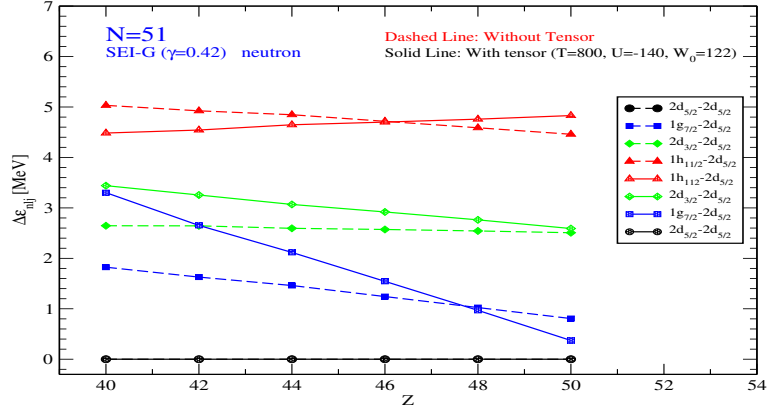}
		\caption{Neutron s.p levels in $N$=51 isotones relative to $2d_{5/2}$ with and without Tensor.}
		\label{N51_n}
	\end{center}
\end{figure}
 \begin{figure}[h]
 	\begin{center}
 		\includegraphics[height=7cm,width=12cm]{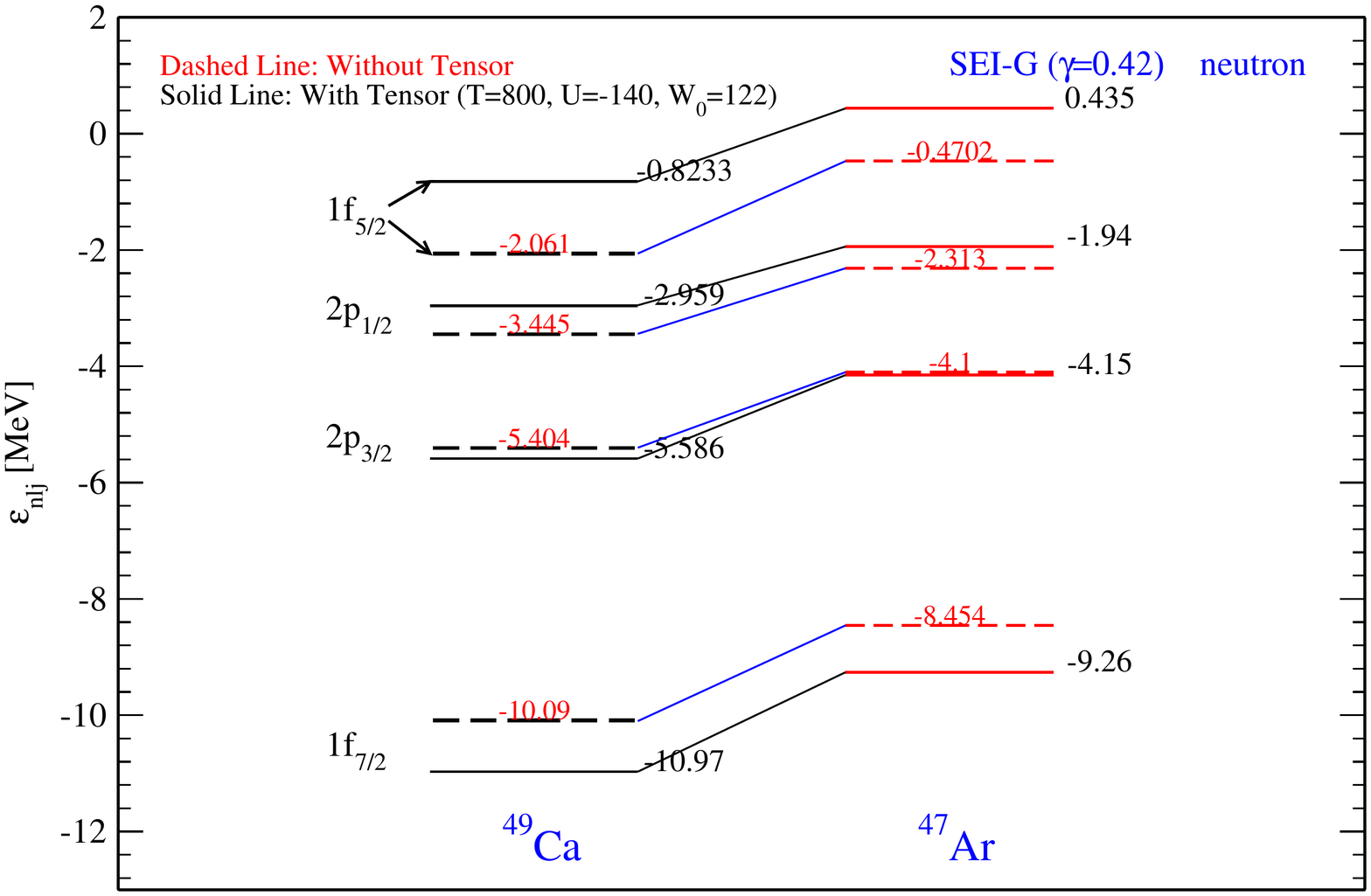}
 		\caption{Neutron single-particle energies (SPE) of the fp orbitals for the $^{47}Ar_{29}$ and $^{49}Ca_ {29}$ 	                    nuclei with and without Tensor}
 		\label{Ca_Ar}
 	\end{center}
 \end{figure}
The SEI model plus the tensor force is also able to reproduce the experimental trends of evolution of the $1h_{11/2}$, 
$1g_{7/2}$ and $2d_{3/2}$ neutron s.p. levels in the $N$=51 isotonic chain. As $Z$ increases from 40 to 50 filling the $1g_{9/2}$ proton level, the  
$1h_{11/2}$ neutron s.p. level is pushed up while the 1$g_{7/2}$ and the $2d_{3/2}$ s.p. neutron levels are pulled down owing to the tensor 
interaction. The lowering of the 1$g_{7/2}$ neutron level is a phenomenon pointed out by Federman and Pittel \cite{Federman1977}.
These effects are shown in Fig.\ref{N51_n}, where the evolution of these s.p. neutron levels relative to the $2d_{5/2}$ neutron level are shown   
for the cases of without and with tensor force. From this figure we can see that the lowering of 1$g_{7/2}$ along this chain is more prominent than 
the one experienced by the $2d_{3/2}$ level, owing to the larger overlap of the wave function of the $1g_{9/2}$ proton level with the one of the 
$1g_{7/2}$ neutron level as compared with the overlap with the wavefunction of the $2d_{3/2}$ level.

\subsubsection*{The $N$=28 gap}

For the nucleus $^{47}$Ar, the SEI model with $\gamma$=0.42 predicts the binding energy and spin-parity of the ground-state as well as the ordering
of the first excited levels, assumed of s.p. nature, in agreement with the experimental values \cite{Gaudefroy2006}.
Although this model also predicts a reduction of 332 keV of the $N$=28 gap in passing from $^{49}$Ca to $^{47}$Ar \cite{Gaudefroy2006},
it fails in the estimation of the reduction of the splitting of the $f$ and $p$ s.p. levels, which are predicted by SEI to be 45 KeV and 172 KeV,
respectively, in comparison with the experimental values of 875 KeV and 890 KeV  \cite{Gaudefroy2006}. This failure can be cured, at least partially,
by including a tensor component in the effective interaction. The neutron s.p. energies in the nuclei $^{49}$Ca and $^{47}$Ar are shown in
Figure \ref{Ca_Ar}. We can see that the introduction of a short-range tensor term with T=800 MeV and U=-140 MeV pulls down the 1$f_{7/2}$ and
2$p_{3/2}$ neutron levels and push up the 1$f_{5/2}$ and 2$p_{1/2}$ ones, due to the monopole effect of the interaction of neutrons
with protons in the occupied level 1$d_{3/2}$ \cite{Otsuka2005}. As a consequence of the tensor contribution, the $N$=28 gap decreases slightly
to 274 keV, but the splittings of the 1$f$ and 2$p$ levels in $^{47}$Ar increases noticeably up to 451 and 417 KeV,
              %
%
        	

\section*{Conclusion}

In this work we have used the SEI model to examine the influence of it's mean-field properties, in particular the NM incompressibility $K$($\rho_{0}$), on the 
inversion of the spin-parity of the ground state observed in $Cu$-isotopes, which has a direct impact on the crossing of the $2p_{3/2}$ and $1f_{5/2}$ 
s.p. states in the $Ni$-isotopes. For this purpose we have considered four EoSs of the SEI family having $\gamma$=$\frac{1}{6}$, $\frac{1}{3}$, 
$\frac{1}{2}$ and $\frac{2}{3}$, which correspond to $K$($\rho_{0}$)=207, 226, 245 and 263 MeV, respectively. The mean-field calculation have been 
performed within the QLDFT framework using the uniform blocking method to describe odd nuclei. This approximation is justified, as far as we have shown in 
a recent work \cite{Routray2021}, QLDFT and HF calculations give identical results in $Ni$-isotopic series. It is shown 
that in case of the SEI interaction with an incompressibility value $K$($\rho_{0}$)=240 MeV, there is no requirement of explicit inclusion of a tensor 
part, unlike the case in Gogny and Skyrme forces, to achieve the $2p_{3/2}$ and $1f_{5/2}$ s.p. level crossing in $Ni$-isotopes (or the spin-parity 
inversion of the ground-state of $Cu$-isotopes) at the mass number $A$=75, which is the value extracted from the experiment \cite{Olivier,Sahin}. 
For SEI EoSs corresponding to $K$($\rho_{0}$) value greater than 240 MeV, the crossing of these s.p. levels occur for an isotope having $N<$46 and 
vice-versa. 
We also show that the spin-orbit interaction in case of SEI does not turn to be highly repulsive, as observed in case of Gogny D1M case, 
displaying a smoothly varying behaviour as the neutron number increases from 40 to 50 along the $Ni$ isotopic chain. This behaviour is different from
the one exhibit by the Gogny D1M force, which shows a high repulsion at small distances. This repulsive behaviour is moderated by inclusion of a 
tensor force as a consequence of which the crossing of the s.p.levels in the $Ni$ isotopic chain occurs at the right mass number. \\
We have also examined the influence of the incompressibility of the SEI EoS on the $2s_{1/2}$ and $1d_{3/2}$ proton s.p. level inversion in $Ca$ 
isotopic chain. It is found that the EoS having $K(\rho_0)$=240 MeV ($\gamma$=0.42) reproduces the inversion effects in the $Ca$ isotopic chain 
closely to the experimental data.\\
We have analyzed other scenarios where the effect of the tensor force is relevant for SEI to explain the experimental trend. To perform this study 
we have added to our SEI a short-range 
tensor interaction with two open parameters. 
Using SEI plus the zero-range tensor force we have studied the the gaps between the $1h_{11/2}-1g_{7/2}$ proton s.p.levels in $Sn$ isotopes and 
the $1i_{13/2}-1h_{9/2}$ s.p. neutron level gaps in the $N$=82 isotonic chain comparing to the available data 
reported in Ref.~\cite{Schiffer2004}. We have also analyzed the action of the tensor force on the evolution of the $1h_{11/2}$, $1g_{7/2}$ and 
$2d_{3/2}$ neutron s.p. levels in the $N$=51 isotonic chain. The tensor force produces a lowering of the  $1g_{7/2}$ neutron levels when 
the occupancy of the $1g_{9/2}$ proton level grows, in agreement with the findings of Federman and Pittel \cite{Federman1977}.  
These examples show that the SEI predictions qualitatively reproduce the experimental trends along the considered 
isotopic and isotonic chains, while the SEI without the tensor can not predict the experimental trend. 
We have also studied the variation of the $2p$ and $1f$ level splittings as well as the energy gap at $N$=28 in passing from $^{49}$Ca to $^{47}$Ar, 
for which experimental information are available \cite{Gaudefroy2006}. We find that SEI can reproduce the gap of 330 MeV at $N$=28 but predict a 
small variation of the $2p$ and $1f$ splittings in passing from $^{49}$Ca  to $^{47}$Ar. This can be cured, at least partially, by the inclusion of 
the tensor interaction, which in spite of a slight reduction of the energy gap at $N$=28, increases considerably the difference of the $2p$ and $1f$
splittings.
In this work we see that by including a short-range tensor term to the standard spin-orbit interaction one is able to explain in a 
qualitative way the experimentally observed mass dependence of the specific energy gaps in $Sn$- and $Sb$-isotopes, $N$=82 and $N$=51 isotonic 
chains as 
well as the reduction in the energy gap at $N$=28 qualitatively. But to have more quantitative explanation, it appears that the tensor and the 
spin-orbit interactions should be modified, for example by introducing finite range in the tensor force and by exploring a more flexible spin-orbit 
part, which are tasks for a future research.
\section*{Acknowledgement}
P.Bano acknowledges the support from MANF Fellowship of UGC, India. TRR acknowledges sincere thanks to Prof. B. Behera for 
meaningful discussions.
M.C. and X.V. were partially supported by Grants No.\ FIS2017-87534-P from MINECO
and No.\ CEX2019-000918-M from  AEI-MICINN through the ``Unit of Excellence Mar\'{\i}a de Maeztu
2020-2023'' award to ICCUB. The work of L.M.R. was partly supported by the Spanish
MINECO Grant No.\ PGC2018-094583-B-I00. M.A has been partially supported by the Spanish MINECO Grant No.\ PID2019-104888GB-I00.

\end{document}